\documentclass[12pt]{article}
 \usepackage{amsmath,amsfonts, amssymb}
\usepackage{hyperref}
\usepackage{graphicx}
\usepackage{psfrag}

\newcommand{\ra}{\rightarrow}

\newcommand{\be}{\begin{equation}}
\newcommand{\ee}{\end{equation}}
\newcommand{\ba}{\begin{eqnarray}}
\newcommand{\ea}{\end{eqnarray}}
\newcommand{\bi}{\begin{itemize}}
\newcommand{\ei}{\end{itemize}}

\newcommand{\Tr}{{\rm Tr}}

\newcommand{\R}{\mathbb{R}}
\newcommand{\C}{{\mathbb C}}
\newcommand{\p}{\partial}

\newcommand{\Ncal}{{\mathcal N}}

\newcommand{\Ocal}{{\mathcal O}}
\newcommand{\Ecal}{{\mathcal E}}

\newcommand{\nn}{\nonumber}
\newcommand{\mo}{{-1}} 

\newcommand{\f}{\frac}
\newcommand{\half}{\frac{1}{2}}
\newcommand{\oo}{\frac{1}}

\newcommand{\aslash}[1]{\,\,{\raise.15ex\hbox{/}\mkern-12mu #1}}
\newcommand{\bslash}[1]{\,\,{\raise.15ex\hbox{/}\mkern-9mu #1}}

\renewcommand{\bar}{\overline}
\renewcommand{\tilde}{\widetilde}
\renewcommand{\hat}{\widehat}

\renewcommand{\title}[1]{\vbox{\center\LARGE{#1}}\vspace{5mm}}
\renewcommand{\author}[1]{\vbox{\center#1}\vspace{5mm}}
\newcommand{\address}[1]{\vbox{\center\em#1}}
\newcommand{\email}[1]{\vbox{\center\tt#1}\vspace{5mm}}

\begin{document}
\bibliographystyle{utphys}  

\begin{titlepage}

\begin{center}
\hfill {\tt NSF-KITP-08-96}\\

\bigskip\bigskip

\title{ Wilson loop correlators at strong coupling:\\
from matrices to bubbling geometries}

\author{Jaume Gomis$^{1,a}$,
Shunji Matsuura$^{1,2,b}$,
Takuya Okuda$^{3,c}$,
and Diego Trancanelli$^{4,d}$}

\address{$^1$ Perimeter Institute for Theoretical Physics\\
Waterloo, Ontario, N2L 2Y5, Canada\\
\medskip
$^2$  Department of Physics, University of Tokyo\\ 7-3-1 Hongo, Bunkyoku, Tokyo 
113-0033, Japan \\
\medskip
$^3$ Kavli Institute for Theoretical Physics, University of California\\
Santa Barbara, CA 93106-9530, USA\\
\medskip
$^4$ Department of Physics, University of California,\\
Santa Barbara, CA 93106-9530, USA}

\medskip

\email{$^a$ jgomis@perimeterinstitute.ca,
$^b$ smatsuura@perimeterinstitute.ca,\\
$^c$ takuya@kitp.ucsb.edu,
$^d$ dtrancan@physics.ucsb.edu}

\end{center}

\abstract{
\noindent
We compute at strong coupling the large $N$ correlation functions of  supersymmetric Wilson loops in large representations of the gauge group with local operators of ${\cal N}=4$ super Yang-Mills.  The gauge theory computation of  these correlators  is performed using matrix model techniques. We show that the strong coupling correlator of the Wilson loop with the stress tensor computed using the matrix model exactly matches the  semiclassical computation of the correlator of the 't~Hooft loop  with the stress tensor, providing a non-trivial quantitative  test of electric-magnetic duality of ${\cal N}=4$ super Yang-Mills. We then perform these calculations using the dual bulk gravitational picture, where the Wilson loop is described by a ``bubbling'' geometry. By applying  holographic methods to these backgrounds we calculate  the Wilson loop correlation functions, finding perfect agreement with our gauge theory results. 
 }

\vfill

\end{titlepage}

\tableofcontents


\section{Introduction and outline}

Typically, computations in four-dimensional gauge theories can only be carried out in the  weak coupling regime, 
where a wealth of perturbative techniques have been developed. 
Dualities in field theory, however,  provide new avenues in which to study the strong coupling behavior of certain field theories by mapping the strong coupling dynamics of one theory  to the weakly coupled regime of the dual theory.
Moreover, some gauge theories are holographically dual to quantum gravity with certain asymptotics, and the strong coupling dynamics of the gauge theory can be solved in terms of semiclassical gravitational physics.

The best understood example, ${\cal N}=4$ super Yang-Mills, is a field theory that both presents electric-magnetic duality \cite{Montonen:1977sn, Witten:1978mh, Osborn:1979tq} and  describes  holographically  quantum gravity with $AdS_5\times S^5$ boundary conditions \cite{Maldacena:1997re}. Due to its high degree of symmetry, it enjoys remarkable properties in the large $N$ limit -- such as integrability -- that allow for the study of some questions in the strong coupling regime (see for instance \cite{Beisert:2006ez}).

In this paper we compute the large $N$, strong coupling correlation functions of a supersymmetric Wilson loop in a large representation of the gauge group\footnote{This is a representation where the number of boxes in each row or column of the corresponding Young tableau is of order $N$.} with local operators of ${\cal N}=4$ super-Yang Mills, specifically with chiral primary operators and the stress tensor. 
We compute these correlators both in   gauge theory and using the dual supergravity description. In   gauge theory we obtain strong coupling results by solving the normal matrix model that captures these correlation functions. We then perform a quantitative test of S-duality of ${\cal N}=4$ super Yang-Mills by also calculating in the semiclassical approximation the correlator between a 't Hooft loop operator and the same local operator. We find that the S-dual of the semiclassical 't Hooft loop correlator exactly matches the strong coupling result of the Wilson loop correlator, providing a quantitative test of electric-magnetic duality in ${\cal N}=4$ super Yang-Mills. We also perform the calculation of the Wilson loop correlation functions using ``bubbling'' geometries and find exact agreement with the strong coupling results we obtained in the gauge theory.

The study of supersymmetric Wilson loops in the context of the AdS/CFT correspondence \cite{Maldacena:1997re,Gubser:1998bc,Witten:1998qj} is important for several reasons. Among them is the fact that these operators  couple to strings  and branes in the bulk, thus touching on stringy properties of the theory. Moreover, Wilson loops allow in some cases, for instance when they follow circular contours,  to obtain results that are exact  in $N$ and the 't~Hooft coupling $\lambda\equiv g^2_{YM}N$. The first example of an all order computation in $N$ and $\lambda$ was the computation of the expectation value of the circular Wilson loop in the fundamental representation, which was conjectured in \cite{Erickson:2000af,Drukker:2000rr} to be captured by a hermitian matrix model. This result has recently been proven using localization techniques 
\cite{Pestun:2007rz} and generalizations thereof have been found in, {\it e.g.}, \cite{Drukker:2006ga,Semenoff:2006am,Drukker:2007dw,Drukker:2007yx,Drukker:2007qr}.

It is a well-established entry in the AdS/CFT dictionary that a supersymmetric Wilson loop in the fundamental representation corresponds in the bulk to a classical string surface with $AdS_2$ induced metric, which extends in the interior of the $AdS$ space and lands on the loop on the boundary \cite{Rey:1998ik,Maldacena:1998im}. In particular, the expectation value of the loop operator is given by the minimal area of the string world-sheet, upon the appropriate regularization of the IR divergence associated with the infinite area of the string \cite{Drukker:1999zq}.

When one considers Wilson loop operators in representations higher than the fundamental, with rank of order ${\cal O}(N)$, the string in the bulk gets replaced by configurations of probe branes with electric flux dissolved in their world-volumes \cite{Gomis:2006sb} (see also  \cite{Drukker:2005kx, Yamaguchi:2006tq, Gomis:2006im, Drukker:2006zk}).
 More specifically, a Wilson loop in the rank $k$ symmetric representation is described by a D3-brane with $k$ units of flux and wrapping an $AdS_2 \times S^2$ subspace \cite{Drukker:2005kx, Gomis:2006im}, whereas a loop  in the rank $k$ antisymmetric representation is described by a D5-brane \cite{Gomis:2006sb, Yamaguchi:2006tq}, also with $k$ units of flux, wrapping an $AdS_2\times S^4$ subspace. These branes can be thought of as emerging from $k$ coincident strings via the Myers polarization effect \cite{Myers:1999ps}, which, for large enough $k$, blows up an $S^2$ or an $S^4$ from the world-sheet  of the coincident strings. 
\begin{figure}[tb]
\begin{center}
\psfrag{n1}{$n_1$}
\psfrag{K1}{$K_1$}
\psfrag{n2}{$n_2$}
\psfrag{K2}{$K_2$}
\psfrag{Kg-1}{$K_{g-1}$}
\psfrag{ng}{$n_g$}
\psfrag{Kg}{$K_g$}
\psfrag{ng+1}{$n_{g+1}$}
\includegraphics[width=80mm]{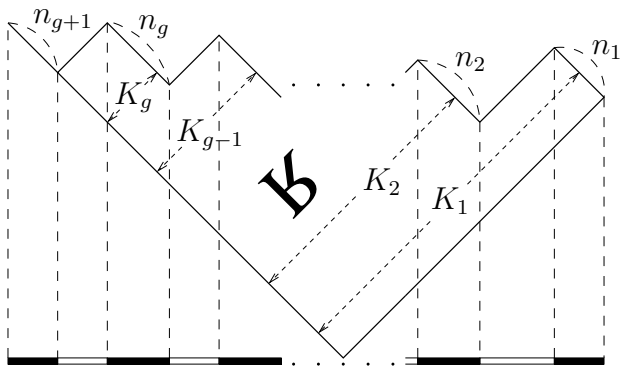} 
\caption{We depict here, rotated and inverted, the Young tableau $R$ of the irreducible representation of $U(N)$   in which we take the Wilson loop operator. 
The tableau consists of $g$ blocks, the $I$-th one of them having $n_I$ rows of length $K_I$. All the edges of the diagram are taken to be long, meaning that $n_I$ and $K_I$ are both of order ${\cal O}(N)$ for all $I$. This guarantees that the dual bubbling geometry has small curvature everywhere.
}
\label{parametrization}
\end{center}
\end{figure}

This probe approximation breaks down when the representation $R$ of the Wilson loop is taken to be even larger, with a corresponding Young tableau containing order ${\cal O}(N^2)$ boxes (see Figure \ref{parametrization}). In this case the  back-reaction of the brane configuration in \cite{Gomis:2006sb} cannot be ignored and the space-time is deformed into a new geometry containing bubbles of new cycles carrying fluxes, and it is thus called a {\it bubbling} solution. The study of the bubbling solutions for  this case was initiated in  \cite{Yamaguchi:2006te, Lunin:2006xr} (see also \cite{Gomis:2006cu}) and culminated  in \cite{D'Hoker:2007fq}, where an elegant description of the solution was given in terms of a Riemann surface. If the representation of the Wilson loop is large, then the dual geometry is guaranteed to have small curvature everywhere. The matrix model prediction for the on-shell action of these bubbling solutions was found in \cite{Okuda:2007kh}.

This entire picture of strings metamorphosing into branes and branes undergoing geometric transitions into new geometries is remarkably universal in the context of gauge theory/gravity dualities, having a very close analog for local operators, where these phenomena were in fact analyzed first \cite{McGreevy:2000cw,Lin:2004nb},\footnote{The bubbling construction for chiral primary operators was found building on ideas introduced in \cite{Berenstein:2004kk}.} and for topological theories \cite{Gomis:2006mv,Gomis:2007kz,Halmagyi:2007rw}.  

The aim of this paper is to deepen our understanding of Wilson loops in large representations of the gauge group, both from a gauge theory  perspective and in the bulk. The tool we use to do this is the detailed study of the correlation functions of these loops with local operators of ${\cal N}=4$ super Yang-Mills, such as chiral primary operators and the stress tensor. We perform computations both in field theory and in supergravity.

In the gauge theory, the computation of these correlation functions makes use of a matrix model (the computation of  the correlators when the Wilson loop is in the fundamental, symmetric and antisymmetric  representation  was performed in \cite{Semenoff:2001xp,Giombi:2006de}).\footnote{For the corresponding Wilson loop expectation value computation see \cite{Okuyama:2006jc,Hartnoll:2006is,Okuda:2007kh}.} Building on the results in \cite{Okuda:2008px}, we solve the matrix model capturing these correlation functions 
in the strong coupling regime and    when the Wilson loop is in a large representation. 

On the other hand, in the bulk analysis we use holographic methods on the bubbling supergravity backgrounds  to extract the desired correlation functions, finding perfect agreement with our  computations in gauge theory. The bubbling supergravity backgrounds indeed contain non-trivial dynamical information about correlation functions (see
\cite{Skenderis:2007yb, Drukker:2008wr} for the computation of correlation functions  from bubbling geometries for   local operators and surface operators respectively).\footnote{See \cite{Berenstein:1998ij, Giombi:2006de} for the probe string/brane computation of the correlation functions when the Wilson loop is in the fundamental, symmetric and antisymmetric representation.}

These Wilson loops in large representations and their dual bubbling geometries represent an arena with an incredibly rich structure, yet one where explicit computations and highly non-trivial quantitative tests of the AdS/CFT correspondence are possible and where one may be able to shed new light into the inner workings of holography.


\subsection{Outline of the paper}

In Section \ref{sec-sym}, we establish the notation and analyze the general structure of the correlation functions we are interested in. The position dependence, both for correlators with chiral primary operators \cite{Semenoff:2001xp} and correlators with the stress tensor \cite{Kapustin:2005py}, is completely determined by the symmetry of the system, so that the computation boils down to finding  coefficients which depend on the characteristic data of the Wilson loop and the local operator, as well as $\lambda$ and $N$. We show using supersymmetric Ward identities that the correlator of the Wilson loop with the stress tensor can be obtained from the correlator of the Wilson loop with the dimension two chiral primary operator. In Appendix  \ref{GL-sec} we derive the same relation using a topological field theory argument.

In Section \ref{CFT-sec}, we proceed to compute  in gauge theory the correlation  coefficients mentioned above. The important point to stress is that we manage to perform these computations at strong coupling, making it possible to compare and match them with the results in supergravity of Section \ref{sugra-sec}.

More specifically, we begin in Section \ref{CPO-sec} with the computation of the correlator between a half-BPS circular Wilson loop and  a chiral primary operator of  ${\cal N}=4$ super Yang-Mills as well as with the stress tensor. 
Similarly to what happens for the expectation values, it has been conjectured in \cite{Semenoff:2001xp} that the exact path integral describing this correlation function is also captured by a matrix model, which sums all the ladder/rainbow diagrams in the perturbative expansion and therefore allows to extract its strong coupling behavior (a derivation using 
localization along the lines of  \cite{Pestun:2007rz} should also be possible). The particular matrix model we use is the {\it normal matrix model} introduced in this context in \cite{Okuyama:2006jc}. Using results obtained in  \cite{Okuda:2008px}, we solve the model for large 't Hooft coupling and large representations of the gauge group and find the moments of the eigenvalue distribution, in terms of which the correlators are determined.

We then calculate in Section \ref{stress-sec} the strong coupling correlator between the same half-BPS circular Wilson loop and the stress tensor of ${\cal N}=4$ super Yang-Mills using a semiclassical computation of the correlator of the 't Hooft loop with the stress tensor. 
This consists in  computing first  the correlator of a half-BPS 't Hooft loop operator with   the stress tensor  in the semiclassical  approximation. To obtain the strong coupling result for the Wilson loop we act with  S-duality on the 't Hooft loop semiclassical  result. This computation yields precisely the same answer computed  by the matrix model in section  \ref{CPO-sec}  for a specific choice of representation of the Wilson loop. We comment on the reason why this happens. This yields a quantitative test of S-duality in ${\cal N}=4$ super Yang-Mills.

The supergravity analysis of these same correlators is contained in Section \ref{sugra-sec}. First, we briefly review the bubbling solution found for the first time in closed form in \cite{D'Hoker:2007fq} and re-express it in terms of the resolvent of the matrix model, which encodes the correlation functions in the gauge theory. We then apply the Kaluza-Klein holography machinery \cite{Skenderis:2006uy,Skenderis:2007yb} to this geometry and extract from the asymptotic expansion of the supergravity fields the correlation functions with chiral primary operators and the stress tensor (see also \cite{Skenderis:2007yb, Drukker:2008wr}). Differently from the gauge theory computation, we are able to carry on the bulk computation only for operators up to dimension four, albeit the procedure we use is in principle applicable to operators of arbitrarily high dimension. The correlators that we compute in supergravity are found in perfect agreement with the strong coupling gauge theory results. Particularly remarkable is the agreement between the correlators with dimension four operators (both the dimension four chiral primary operator and the stress tensor), because of very delicate cancellations between non-linear terms that take place in supergravity, as expected from the strong coupling gauge theory analysis.

We conclude the paper with a series of appendices in which we collect some technical details of our calculations.


\section{Symmetry analysis of the Wilson loop correlators}
\label{sec-sym}

In this paper we study the half-BPS circular Wilson loop of  ${\cal N}=4$ super Yang-Mills in $\mathbb{R}^4$. It is given by
\ba
W_R(\theta,a)\equiv\oo{\dim R} \Tr_R {\rm P} \exp\oint_{\rm circle}
 \left(i A+ \phi^i\theta^i |\dot x|ds\right)\, . 
\label{W-def}
\ea
The trace is taken over an irreducible representation $R$ of $U(N)$.
The Wilson loop $W_R$, besides containing the holonomy of the gauge field $A=A_\mu dx^\mu$, also couples to the scalars $\phi^i$ of the ${\cal N}=4$ multiplet through $\theta^i$, a constant unit vector on $\mathbb{R}^6$. The integral is taken along a circle of radius $a$ in $\mathbb{R}^4$ parametrized by $0\leq s\leq 2\pi$.

The circular Wilson loop is related to the
straight Wilson line
\ba
W_R^{\rm line}(\theta)=\oo{\dim R}
\Tr_R {\rm P} \exp\int_{\rm line} 
\left(i A+ \phi^i\theta^i |\dot x|ds\right)\,  
\label{line}
\ea
by a conformal transformation (an inversion around the origin).
Despite this relation, the expectation value of the straight Wilson line
is trivial (independent of the 't Hooft coupling $\lambda\equiv g^2_{YM}N$ and $N$),
while the expectation value of the circular Wilson loop has a non-trivial dependence on $\lambda$ and $N$ \cite{Erickson:2000af,Drukker:2000rr}, which can be interpreted as 
a conformal anomaly \cite{Drukker:2000rr}.

 We shall see in the following that it is  also useful to study the physics of the   half-BPS  Wilson loop of  ${\cal N}=4$ super Yang-Mills by considering the theory on $AdS_2\times S^2$ instead of $\mathbb{R}^4$  \cite{Kapustin:2005py}. In this geometry the Wilson loop is inserted along the boundary of $AdS_2$.\footnote{To be precise, the Wilson loop should be inserted on a curve at a finite
distance away from the boundary and then we should take the limit in which this curve approaches the boundary. We refer the reader to \cite{Kapustin:2005py} for details on the procedure.
} The straight Wilson line corresponds to taking  the metric on $AdS_2$ in Poincar\'e coordinates   while the circular Wilson loop corresponds to taking $AdS_2$ in global coordinates.\footnote{Throughout this paper  $AdS$ refers always to Euclidean  $AdS$ space.}
 As we show in Appendix \ref{sec-conformalfactor},
 the metric of $AdS_2\times S^2$ in global and Poincar\'e coordinates is related to the metric in $\mathbb{R}^4$  
by a Weyl transformation, which allows us to relate the computations in $\mathbb{R}^4$ to the computations in 
$AdS_2\times S^2$.

We now proceed to study   the correlators of the half-BPS Wilson loop  with chiral primary operators 
and with the stress tensor in  ${\cal N}=4$ super Yang-Mills.


\subsection{Correlators with chiral primary operators}
\label{loop-CPO}

A chiral primary operator of dimension $J$ is a scalar operator  transforming in the representation of the $SO(6)$ R-symmetry with the highest weight  $(0,J,0)$. It is given by\footnote{Here  $\phi_i=\phi_i^aT^a$. The gauge group generators are canonically normalized by $\Tr(T^aT^b)=\f{1}{2}\delta^{ab}$ and
the scalar propagator is given by  $
\langle \phi_i^a(x)\phi_j ^b(0)\rangle_{YM}
=\f{g_{YM}^2}{4\pi^2}\f{\delta_{ij}\delta^{ab}}
{x^2}
$.} 
\ba
{\cal O}_J(x)\equiv \f{(8\pi^2)^{J/2}}{
\lambda^{J/2} \sqrt{J}}C_{i_1\ldots i_J}  \Tr\, \phi^{i_1}\ldots \phi^{i_J}(x)\, ,
\label{unit}
\ea
where $Y=C_{i_1\ldots i_J}\theta^{i_1}\ldots \theta^{i_J}$
is an $SO(6)$ scalar spherical harmonic   that
specifies the operator $\Ocal_J$.
We normalize the complex coefficients $C_{i_1\ldots i_J}$ so that
\ba
C_{i_1\ldots i_J} \bar C_{i_1\ldots i_J}=1,
\ea
corresponding to the normalization of the spherical harmonics $Y$ given by
\ba
\int_{S^5}| Y(\theta)|^2=\f{\pi^3}{2^{J-1}(J+1)(J+2)}\, .
\label{Y-norm}
\ea
The operators (\ref{unit}) are normalized such that 
their two-point
function is unit normalized in the planar approximation
\ba
\langle {\cal O}_J(x){\cal O}_J(y)\rangle=\oo{ |x-y|^{2J}}\, .
\ea

Let us first consider the correlator of the Wilson loop
and $\Ocal_J(x)$ in $AdS_2\times S^2$.
Because this space is homogeneous (all points are related to
each other by an isometry), 
the correlator is independent of
the position $x$.
By taking into account the transformation properties  under the R-symmetry group,
we can parametrize the correlator as
\ba
\langle \Ocal_J(x)\rangle_{W}=\Xi_{R,J}Y(\theta).
\label{WO-AdS}
\ea
The notation $\langle ...\rangle_W$ denotes
a  correlator of ${\cal N}=4$ super Yang-Mills in $AdS_2\times S^2$ with the Wilson loop $W_R(\theta)$
on the boundary, normalized so that $\langle 1 \rangle_W=1$. This expression holds
both when $AdS_2$  is in global and Poincar\'e coordinates, corresponding to inserting a circular
Wilson loop and straight Wilson line respectively.

For the circular Wilson loop (\ref{W-def}) in $\mathbb{R}^4$,
the correlator is given by
\ba
\f{\langle W_R(\theta,a) 
{\cal O}_J(x)
\rangle}
{\langle W_R(\theta,a)\rangle}
=\Xi_{R,J} Y(\theta)\f{1}{ \tilde{r}^J}\, ,
\label{corr-coef}
\ea
where we have defined  the conformally invariant distance  $\tilde r$ as
\ba
\tilde r=\f{\sqrt{(r^2+L^2-a^2)^2+4a^2L^2}}{2a}\, ,
\label{tilda}
\ea
which is also the conformal factor that relates the metric on $\R^4$ to the metric on $AdS_2\times S^2$ in global coordinates,
as we show in Appendix \ref{sec-conformalfactor}.
Here $a$ is the radius of the $S^1$, $L$ is the distance between the location  of the chiral primary operator (\ref{unit}) 
and the plane that contains the circle.
The other parameter $r$ is the distance between the location of the operator (\ref{unit}) 
and the axis of the circle.
Therefore, 
symmetries determine the correlator  between a chiral primary operator and a circular Wilson loop   up to the coefficient  $\Xi_{R,J}$. 

Similarly, the correlator of the straight Wilson line in $\mathbb{R}^4$  (\ref{line})
and the chiral primary (\ref{unit}) is given by
\ba
\langle W^{\rm line}_R(\theta)
{\cal O}_J(x)
\rangle
=\Xi_{R,J} Y(\theta)\oo{l^J}\, ,
\label{corr-coef-line}
\ea
where $l$ can be interpreted, again, both as the distance between the line and ${\cal O}_J(x)$
as well as the conformal factor relating the metric on  $\R^4$ to the metric on $AdS_2\times S^2$ in Poincar\'e coordinates (see Appendix \ref{sec-conformalfactor}).

Agreement between various computations we perform in this paper
suggests that the correlator of a Wilson loop with a local operator
normalized by the Wilson loop expectation value transforms simply under  conformal
transformations, so that $\Xi_{R,J}$ is the same for the correlator computed with the circular Wilson loop 
and with the straight Wilson line. 
The transformation properties of this ratio seem to be solely determined by the representation of the local operator under the conformal group, and does not suffer from the conformal anomaly of the  Wilson loop expectation value 
\cite{Drukker:2000rr}. 
It would be interesting to prove this lack of conformal anomaly of the normalized correlator from first principles. 
 
The coefficient $\Xi_{R,J}$ 
depends on $R$, the representation of the Wilson loop, and $J$, the dimension of the chiral primary operator (it is also a non-trivial function of the 't Hooft coupling $\lambda$ and $N$) but is independent of the choice of  operator in the $(0,J,0)$ multiplet of $SO(6)$ one uses. We will take advantage of this by choosing convenient operators in the multiplet for the various computations.


\subsection{Correlator  with the stress tensor}

The correlator of a half-BPS Wilson loop with the stress tensor is also essentially fixed by symmetries up to a scaling function $h_W$, that we wish to compute. For the stress tensor one must take into account that $U(N)$ ${\cal N}=4$ super Yang-Mills on a curved background has a conformal anomaly given by
\ba
\langle T_\mu^{\;\mu}\rangle=\f{N^2}{32\pi^2}\left(R_{\mu\nu}R^{\mu\nu}-\f{1}3 R^2\right),  
\label{anom}
\ea
where $R_{\mu\nu}$ is the Ricci tensor and $R$ is the Ricci scalar of the background.
This relation is protected from quantum corrections 
because the superconformal symmetry relates
the stress tensor to derivatives of the R-current.

On   the  $AdS_2\times S^2$ background we are considering (both in global and Poincar\'e coordinates),   the anomaly (\ref{anom}) is $N^2/8\pi^2$.
This, together with the symmetries of the problem, determines  the correlator up
to a real number $h_W$ -- the scaling function -- that depends on $g_{YM}, N$, and $R$,
but not on $\theta$ \cite{Kapustin:2005py} 
\ba
\langle T_{\mu\nu}(x)\rangle_W dx^\mu dx^\nu
=h_W(ds^2_{AdS_2}-ds^2_{S^2})+ \f{N^2}{32\pi^2}(ds^2_{AdS_2}+ds^2_{S^2}).
\label{T-ads}
\ea

We now turn to ${\cal N}=4$ super Yang-Mills in $\R^4$. In this case there is no conformal anomaly.
The  correlator of the straight Wilson line with  the stress tensor in $\R^4$
is given by  \cite{Kapustin:2005py}
\ba
&&
\langle W_R^{\rm line}(\theta) T_{44}(x)\rangle
=\f{h_W}{l^4}\,,
\qquad 
\langle W_R^{\rm line}(\theta) T_{4a}(x)\rangle
=0\, , 
\nn\\
&& 
\langle W_R^{\rm line}(\theta) T_{ab}(x)\rangle
=-h_W\f{\delta_{ab}-2n_a n_b}{l^4}\, ,
\label{T-line}
\ea
where we have taken the  line along the 4-direction 
and  $n^a=x^a/l$ for $a,b=1,2,3$ is the unit normal vector to the line ({\it i.e.} $n^an^a=1$).

The form of the correlator between the circular loop and the stress tensor in $\R^4$
can be obtained similarly and we write it for completeness. The circular loop in the coordinate system
\be
ds^2_{\mathbb{R}^4}=dr^2+r^2d\psi^2+dL^2+L^2 d\phi^2
\ee
is
supported at $r=a$ and $L=0$. The correlator is then given by
 \ba
&&
\f{\langle W_R(\theta,a) 
T_{rr}(x)
\rangle}
{\langle W_R(\theta,a)\rangle}
= {h_W}\left(\oo{\tilde{r}^4}-\f{2r^2L^2}{a^2\tilde{r}^6}\right)\,,
\qquad 
\f{\langle W_R(\theta,a) 
T_{\psi\psi}(x)
\rangle}
{\langle W_R(\theta,a)\rangle}
= h_W\f{r^2}{\tilde{r}^4}\, , 
\nn\\
&& 
\f{\langle W_R(\theta,a) 
T_{LL}(x)
\rangle}
{\langle W_R(\theta,a)\rangle}
={h_W}\left(\oo{ \tilde{r}^4}-\f{(a^2+L^2-r^2)^2}{2a^2\tilde{r}^6}\right)\,,
~~
\f{\langle W_R(\theta,a) 
T_{\phi\phi}(x)
\rangle}
{\langle W_R(\theta,a)\rangle}=-h_W\f{L^2}{\tilde{r}^4}\, ,
\nn\\
 &&
\f{\langle W_R(\theta,a) 
T_{rL}(x)
\rangle}
{\langle W_R(\theta,a)\rangle}=-{h_W} \f{rL(a^2+L^2-r^2)^2}{a^2\tilde{r}^6}\,.
\label{T-circle}
\ea
It is completely determined by the scaling function $h_W$.


\subsection{Correlator  with the stress tensor from Ward identities}

We now wish to derive an exact relation between the 
Wilson loop correlator with the stress tensor and 
the Wilson loop correlator with the dimension two chiral
primary operator. This will allow us to compute the first correlator from the knowledge of the second one.

The relation between these two correlators is a consequence of a Ward identity. 
The idea is to apply the supersymmetry Ward identity
to operators in the current supermultiplet, which contains both the dimension two chiral 
primary operator and the stress tensor \cite{Bergshoeff:1980is}.

Under the Poincar\'e supersymmetry transformations of ${\cal N}=4$ super Yang-Mills\footnote{We do not write the four-dimensional spinor indices for clarity.}
\ba
\delta \varphi^{AB}&=&\lambda^{[A}\eta^{B]}+\half \epsilon^{ABCD}\bar\eta_C
\bar\lambda_D,
\nn\\
\delta A_\mu &=& -i (\lambda^A\sigma_\mu\bar\eta_A+\eta^A\sigma_\mu
\bar\lambda_A),
\ea
the straight Wilson line  
\ba
W_R^{\rm line}(\theta)=\f{1}{\mbox{dim}R}\Tr_R {\rm P} \exp i \int ds(A_4+\theta^i \Sigma^i_{AB}\varphi^{AB})
\ea
is invariant under the following supersymmetries
\ba
\bar \eta_A=i\theta^i \Sigma^i_{AB}\bar\sigma_4 \eta^B,
\label{susystraight}
\ea
where $\Sigma^i_{AB},\, \bar\Sigma^{iAB}$ are the six-dimensional chiral sigma matrices satisfying (see {\it e.g.} \cite{Green:2002vf})
\ba
\Sigma^{i}_{AB}\bar\Sigma^{j BC}
+
\Sigma^{j}_{AB}\bar\Sigma^{i BC}
=2\delta^{ij}\delta_A^C,\\
\bar\Sigma^{iAB}=-\f{1}{ 2}\epsilon^{ABCD}\Sigma^i_{CD}, 
\ea
and the scalars in the ${\bf 6}$ of $SU(4)$ are given by $\phi^i=\Sigma^{i}_{AB}\varphi^{AB}$.
One can regard
  $\Sigma^{i}_{AB}$ as the Clebsch-Gordan coefficients coupling two ${\bf {4}}$'s to a $\bf{6}$ of $SU(4)$.

Let's now consider the following correlator
\ba
\langle W_R^{\rm line}(\theta)\delta \Ocal(x)\rangle,
\ea
where $\delta$ denotes a supersymmetry transformation generated by the supersymmetries preserved by the Wilson line (\ref{susystraight}) and $\Ocal(x)$ is an arbitrary local operator. Since $\delta W_R^{\rm line}(\theta)=0$ we have that 
\ba
\langle W_R^{\rm line}(\theta)\delta \Ocal(x)\rangle
=
\langle\delta\left( W_R^{\rm line}(\theta)
\Ocal(x)\right)
\rangle
=\langle 
[\eta^{i\alpha} Q_{i\alpha} +\bar \eta_i^{\dot\alpha} \bar Q^i_{\dot \alpha}
,
W_R^{\rm line}(\theta)
\Ocal(x)]
\rangle
=0
\label{ward}
\ea
for any local operator $\Ocal(x)$.

The supersymmetry variations of the supercurrent $J_{\mu A}$ (in the ${\bf 4}$ of $SU(4)$)
and the fermionic
operator $\chi^C_{~AB}=-\chi^C_{~BA}$ (in the ${\bf 20}$ of $SU(4)$)
 in the current supermultiplet are given by (see {\it e.g.} \cite{Basu:2004nt})
\ba
\delta J_{\mu A}&=&-\sigma^\nu T_{\mu\nu} \bar \eta_A
-2(\sigma_\rho \bar\sigma_{\mu\nu}-\oo 3 \sigma_{\mu\nu}\sigma_\rho)
\p^\nu R^{\rho C}{}_A\bar\eta_C
\nn\\
&&
~~~
-(\sigma_{\rho\sigma}\sigma_{\mu\nu}+\oo 3 \sigma_{\mu\nu} \sigma_{\rho\sigma}
)\epsilon_{ACDE} \p^\nu B^{CD\rho\sigma}\eta^E\, ,
\nn\\
\delta \chi^C_{~AB}&=&
\f{3}4\bigg[
i\epsilon_{ABEF}\sigma^{\mu\nu} B_{\mu\nu}^{CE}\eta^F
+i\epsilon_{ABEF}\Ecal^{EC}\eta^F
\nn\\
&&~~~-\sigma^\mu R^C_{~\mu[A}\bar\eta_{B]}
+2i \sigma^\mu\p_\mu Q^{CD}{}_{AB}\bar \eta_D
\bigg]-{\rm trace}\, ,
\label{susycurrent}
\ea
where $a^C_{AB}-{\rm trace}=a^{C}_{AB}-(1/3)
(\delta^C_A a^D_{DB}-\delta_B^C a^D_{DA})$.
The supersymmetry transformations generate other operators in the current supermultiplet. For example, 
in the right hand side of (\ref{susycurrent}) we get
\ba
Q^{AB}_{~~CD}&=&
\f{1}{4g_{YM}^2}
 \bar\Sigma^{AB}_i \Sigma_{jCD}
\Tr\left(\phi^i\phi^j-\oo 6 \delta^{ij} \phi^k\phi^k\right),
\label{Q-phi}
\ea
which is the dimension two chiral primary operator in the ${\bf 20'}$ of $SU(4)$.
The R-symmetry current $R^\mu{}^A_{~B}$, the scalar operator $\Ecal^{AB}=\Ecal^{BA}$ and the two-form  
 $B_{\mu\nu}^{AB}=-B_{\mu\nu}^{BA}$
 transform in the ${\bf 15}$, ${\bf 10}$, ${\bf 6}$ 
representations of $SU(4)$ respectively.

We can now constrain the correlator of the straight Wilson line $W_R^{\rm line}(\theta)$ with these operators by using the fact that $W_R^{\rm line}(\theta)$ is $SO(5)$ invariant. Since the ${\bf 15}$ and ${\bf 10}$ representations of $SU(4)$ do not contain an $SO(5)$ singlet in the decomposition of $SO(5)\subset SU(4)$ we have that 
\ba
\langle W_R^{\rm line}(\theta)  R^\mu{}^A_{~B}\rangle=0\, ,
~~~~\langle W_R^{\rm line}(\theta)  \Ecal^{AB}\rangle=0\, .
\ea
On the other hand, since ${\bf 6}\rightarrow {\bf 1}\oplus {\bf 5}$ under the decomposition, we have that
\ba
\langle W_R^{\rm line}(\theta) B_{4a}^{AB}\rangle=0,
~~~~
\langle W_R^{\rm line}(\theta) B_{ab}^{AB}\rangle=
b \bar\Sigma^{AB}_i\theta^{i}\f{\epsilon_{abc}n^c}{l^3}\, .
\ea
 Likewise, we have from (\ref{corr-coef-line}) and (\ref{Q-phi}) that 
\ba
\langle W_R^{\rm line}(\theta)  Q^{AB}_{~~CD}\rangle
= \f{\sqrt 2 N}{32\pi^2} \Xi_{R,2}\bar\Sigma^{AB}_i \Sigma_{j C D}
(\theta^i\theta^j-\oo 6 \delta^{ij})\oo{l^2}\, ,
\ea
and, as we have already seen,
\ba
&&
\langle W_R^{\rm line}(\theta) T_{44}(x)\rangle
=\f{h_W}{l^4}\,,
\qquad 
\langle W^{\rm line}_R(\theta) T_{4a}(x)\rangle
=0\, , 
\nn\\
&& 
\langle W^{\rm line}_R(\theta) T_{ab}(x)\rangle
=-h_W\f{\delta_{ab}-2n_a n_b}{l^4}\,.
\ea
These correlators are completely characterized 
  by the  functions $b, \, h_W$  and $\Xi_{R,2}$, that
depend  on the representation $R$, on $g_{YM}^2$  and $N$.

Our goal is to relate these three quantities. For that we use the Ward identity (\ref{ward}) and the supersymmetry transformations (\ref{susycurrent}). 
To do that, 
let us compute
\ba
0&=&\langle W_R^{\rm line}(\theta) \delta J_{4 A}\rangle
\nn\\
&=&
 -\sigma^4 \f{h_W}{l^4} \bar \eta_A
 -(\sigma_{cd}\sigma_{4a}+\oo 3 \sigma_{4a} \sigma_{cd}
 )\epsilon_{ACDE}  b \epsilon_{cde}\p^a\left(\f{n^e}{l^3}\right)
\theta^{i} \bar \Sigma^{CD}
\eta^E
\nn\\
&=&(i h_W-\f{4}3 b) \f{\theta^i \Sigma^i_{AB} \eta^B}{l^4},
\\
0&=&\langle W_R^{\rm line}(\theta) \delta \chi^C_{~AB}\rangle
\nn\\
&=&
\f{3}4\bigg[
i\epsilon_{ABEF}\sigma^{ab} b \bar\Sigma_i^{CE}\theta^i 
\f{\epsilon_{abc}n^c}{l^3}\eta^F
\nn\\
&&~~~
+2i \sigma^a
 \f{\sqrt 2 N}{32\pi^2} \Xi_{R,2}\bar\Sigma^{CD}_i \Sigma_{j AB}
(\theta^i\theta^j-\oo 6 \delta^{ij})
\p_a\left(\oo{l^2}\right)
\bar \eta_D
\bigg]-{\rm trace}
\nn\\
&=&\f{3}4 (b+i \f{\sqrt 2 N}{8\pi^2} \Xi_{R,2}) \f{n^a\sigma^a}{l^3} \theta^i 
(\Sigma_{iAB} \delta^C_D -\oo 3 \delta^C_B \Sigma^i_{AD}+\oo 3
\delta^C_A \Sigma^i_{B D})
\eta^D\,,
\ea
where we have used that the supersymmetry transformation is generated by a spinor satisfying (\ref{susystraight}).
Therefore we obtain that
\ba
h_W=-\f{N}{3\sqrt 2\pi^2}\Xi_{R,2}\, .
\label{relacioentre}
\ea
This relation has been checked at weak coupling to make sure
that the numerical coefficient is correct.
We stress that the relation holds exactly for arbitrary $R, \, g_{YM}^2,$ and $N$, as it follows from a Ward identity.
This allows us to calculate the correlator of the half-BPS Wilson loop with the stress tensor in terms
of the correlator of the Wilson loop with the dimension two chiral primary operator.\footnote{See Appendix \ref{GL-sec} for an alternative 
derivation of this relation obtained using a topological field theory argument based on
the GL twist \cite{Kapustin:2006pk}.} This will allow us to compute the stress tensor correlator at strong coupling by solving a matrix model.

We expect that similar arguments can be constructed
to relate the correlator with  stress tensor to the correlator with the dimension two chiral primary,
in cases involving other supersymmetric operators/backgrounds,
{\it e.g.} surface operators, half-BPS local operators,
and interface CFT's.


\section{Correlators from gauge theory}
\label{CFT-sec}

In this section we compute the coefficients $\Xi_{R,J}$
and $h_W$
in field theory. 
Later, in Section \ref{sugra-sec},  we will repeat these computations in supergravity using the bubbling Wilson loop supergravity solutions.
As explained earlier,  $\Xi_{R,J}$ is independent
of the choice of $C_{i_1\ldots i_J}$,
{\it i.e.}, the choice of the spherical harmonic   $Y(\theta)$, or, equivalently, it is independent of the
choice of operator in the $(0,J,0)$ $SU(4)$ multiplet. We will take advantage of this when calculating $\Xi_{R,J}$ in gauge theory and supergravity.


\subsection{Correlators from a matrix model}
\label{CPO-sec}


\subsubsection{Complex and normal matrix models}
 
So far we have not committed to any explicit choice of chiral primary operator representative in (\ref{unit}). We do this now in order to compute $\Xi_{R,J}$, having in mind that the final result is in fact   independent of this choice. 
We take the following definition of complex scalar field:
\ba
Z\equiv \f{\phi^1+i\phi^2}{\sqrt 2}\, ,
\label{CPO}
\ea
and choose  the following chiral primary operator:   
\ba
{\cal O}_J(x)\equiv {(8\pi^2)^{J/2}\over \lambda^{J/2} \sqrt{J}}\Tr\, Z^J,
\label{CPOunit}
\ea
which, as we shall see shortly, allows for its correlator to be computed 
using a 
matrix model.
The $\theta$ dependence of the correlator of this operator with the circular Wilson loop $W_R(\theta,a)$ is given by (\ref{corr-coef}) with   
\be
Y(\theta)=\f{(\theta^1+i\theta^2)^J}{2^{J/2}}.
\ee

Using the symmetries of the problem, we can specialize without loss of generality to a  configuration with $r=0$, corresponding to the local operator being inserted on the symmetry axis of the circle. We can moreover use the $SO(6)$ symmetry to take $\theta=(1,0,\dots,0)$ so that the Wilson loop (\ref{W-def}) only couples to $\phi^1$. In this case the
contribution to the correlator (\ref{corr-coef}) due to the spherical harmonic associated with (\ref{CPOunit}) is  
 $Y(\theta)=2^{-J/2}$.

It was conjectured in  \cite{Semenoff:2001xp} that
radiative corrections to this correlator
that involve internal vertices cancel to all orders in
perturbation theory and therefore do not contribute to the evaluation of correlators between chiral primaries and circular Wilson loops.
This is the working assumption we make for the gauge theory analysis (a derivation using localization similar to the one in \cite{Pestun:2007rz} should be possible). Moreover,
with the choice $r=0$, every point on the circle is equidistant from $x$ and the propagator between
the chiral primary and the Wilson loop becomes constant.
It was first noticed in \cite{Erickson:2000af} that, in Feynman gauge, the combined propagator 
for the gauge field and the scalars between two points 
on the circle is also a constant (independent of the radius $a$ of the circle).

Summing over all Feynman diagrams reduces then to a combinatorial problem, where one has to count the number of free propagators at any order in perturbation theory.
As pointed out in \cite{Okuyama:2006jc},
this combinatorics
is exactly captured by a complex Gaussian matrix model defined by the partition function ${\cal Z}_C=\int [dz]\exp\left(-\f{2N}{\lambda} \Tr \bar z z\right)$, where $z$ is a complex $N\times N$ matrix. This matrix model also computes the two-point function of local operators in ${\cal N}=4$ super Yang-Mills of the form $\Tr(Z^J)$ \cite{Kristjansen:2002bb}. Therefore, the correlator of the circular Wilson loop 
(\ref{W-def})  with the chiral primary operator (\ref{CPOunit}) is given by 
\ba
\f{\langle W_R(\theta,a) {\cal O}_J(x)\rangle_{YM}}
{\langle W_R(\theta,a)\rangle_{YM}}
={1\over \tilde{r}^J}{1\over \lambda^{J/2} \sqrt{J}}
\f{\int [dz]e^{-\f{2N}{\lambda} \Tr \bar z z}  \Tr_R e^{(z+\bar z)/ 2 }\,
 \Tr z^J}{\int [dz]e^{-\f{2N}{\lambda} \Tr \bar z z}\Tr_R e^{(z+\bar z)/ 2 }},
\label{R4-CM}
\ea
where $\tilde{r}$ is given in (\ref{tilda}).

By comparing this expression with (\ref{corr-coef}), and using that for  $\theta=(1,0,\dots,0)$ the spherical harmonic function
corresponding to (\ref{CPOunit}) is given by $Y(\theta)=2^{-J/2}$, we find that 
\ba
\Xi_{R,J}=
\f{2^{J/2}}{ \lambda^{J/2} \sqrt{J}}
\f{\int [dz]e^{-\f{2N}{\lambda} \Tr \bar z z}  \Tr_R e^{(z+\bar z)/ 2 }\,
 \Tr z^J}{\int [dz]e^{-\f{2N}{\lambda} \Tr \bar z z}\Tr_R e^{(z+\bar z)/ 2 }
}.
\label{Xi}
\ea
Therefore, we arrive at the result that the correlator of a dimension $J$  chiral primary operator $ {\cal O}_J$ with a half-BPS circular Wilson loop $W_R(\theta,a)$ is 
captured by the moment $\langle z^J\rangle$ of a  Gaussian complex matrix model (\ref{R4-CM}).

After having carefully settled the normalization factors, 
the next task is to compute the moments $\langle z^J\rangle$ 
in the complex matrix model.  
This is not easy because the 
eigenvalues of the complex matrix model do not decouple from the off-diagonal components of the matrix. 
To proceed, we will
map the complex matrix model to a {\it normal matrix model}, where one can reduce the computation of the moments to integrals over the eigenvalues. 
Via manipulations involving coherent states, the authors
of \cite{Okuyama:2006jc} proved the following formula  (see Appendix A of \cite{Okuyama:2006jc}):
\ba
&&\oo{{\cal Z}_{H}}\int d^N\xi e^{-\f{2N} \lambda \sum_i \xi_i^2}\Delta(\xi)^2
\prod_i e^{k_i \xi_i}\cr && \hskip 1cm
=
\oo{{\cal Z}_{N}}\int d^{2N}z e^{-\f{2N} \lambda \sum_i \bar z_i z_i}|\Delta(z)|^2
\prod_i e^{k_i \f{z_i+\bar z_i}{\sqrt 2}} e^{-\f{\lambda}{8N}k_i^2}.
\label{gordon}
\ea
On the left hand side we have the hermitian matrix model, with eigenvalues $\xi_i$ and partition function ${\cal Z}_H=\int d^N\xi e^{-\f{2N} \lambda \sum_i \xi_i^2}\Delta(\xi)^2$, while $z_i$ and ${\cal Z}_N=\int d^{2N}z e^{-\f{2N} \lambda \sum_i \bar z_i z_i}|\Delta(z)|^2$ are the eigenvalues and partition function of the normal matrix model. The factors of $\Delta$ are the Vandermonde determinants originating from the transformation to the eigenvalue basis and the constants $k_i$ encode all the information about the representation of the Wilson loop insertion.

Since $\Tr_R (e^\xi)$ is a polynomial of $e^{\xi_i}$,
the equation above proves that the hermitian and the normal matrix model are almost equivalent,
upon the replacement $\Tr_R(e^{\xi}) \ra  \Tr_R(e^{(z+\bar z)/\sqrt 2})$.
Because of the extra factors $e^{-\f{\lambda}{8N}k_i^2}$ in the right hand side,
this equivalence seems limited to the anti-symmetric representations where
these factors are independent of the index $i$ 
and can be pulled out of the integral \cite{Okuyama:2006jc}.
We can circumvent this difficulty by rewriting
\ba
e^{-\f{\lambda}{8N}k_i^2}
=\sqrt{\f{2N}{\pi\lambda}}\int d\alpha_i e^{-\f{2N}\lambda \alpha_i^2+i k_i\alpha_i}.
\label{alpha-trick}
\ea
Thus we find that (going back to the matrix form for conciseness)
\ba
&&\oo{{\cal Z}_{H}}\int [d \xi] e^{-\f{2N} \lambda \Tr \xi^2}
{1\over \dim R}\Tr_R(e^{\xi})
\nn\\&& \hskip 1cm =
\oo{{\cal Z}_{N}{\cal Z}_\alpha}\int_{[z,\bar z]=0} [dz] [d\alpha] e^{-\f{2N} \lambda \Tr( \bar z z+\alpha^2) }
{1\over \dim R}\Tr_R (e^{\f{z+\bar z}{\sqrt 2}+i \alpha})\,,
\label{finalmani}
\ea
where $[d\xi]$ and $[dz]$ are the hermitian and normal matrix measures, respectively,
$\alpha$ is a real diagonal matrix, and $[d\alpha]$ is the Euclidean measure.
We have divided by the appropriate partition functions for proper  normalization.

We are in fact interested in an extended version of the relation 
above, 
which is obtained by applying the trick (\ref{alpha-trick})
to the results 
discussed in   Appendix C of \cite{Okuyama:2006jc}.
This extension includes an extra insertion corresponding to a chiral primary operator,
and is given by
\ba
&&\hskip -1.3cm \oo{{\cal Z}_{H}^2}\int [d\xi][d\eta] 
e^{-\f{2N} \lambda \Tr (\xi^2+\eta^2)}
{1\over \dim R}\Tr_R(e^{\xi}) \Tr (\xi+i\eta)^J
\cr
&& = \f{1}{2^{J/2}}
\f{1}{{\cal Z}_{N}{\cal Z}_\alpha}\int_{[z,\bar z]=0} 
[dz] [d\alpha] e^{-\f{2N} \lambda \Tr (\bar z z+\alpha^2) }
{1\over \dim R}\Tr_R (e^{\f{z+\bar z}{\sqrt 2}+i \alpha})
\Tr z^J\nn\\
&& \hskip 1cm \equiv
 \f{1}{2^{J/2}}  \langle W_R \Tr z^J\rangle_{MM}
\label{normalito}
\ea
This formula rewrites the complex matrix model correlator in (\ref{Xi})
as a normal matrix model correlator. We will calculate $\Xi_{R,J}$ in  (\ref{Xi}) using the normal matrix 
model description.

We are interested in the computation when the representation $R$ is large while 
$J$ is of order one. It is for this class of operators that the dual supergravity background that we will compute
with 
  in the next section has small curvature everywhere. 
In this case, the eigenvalue distribution 
of the hermitian and normal matrix models   is altered  by the   Wilson loop insertion
but not by the chiral primary operator insertion.

The trick  found in \cite{Halmagyi:2007rw} and  used
in \cite{Okuda:2008px}  to analyze 
the eigenvalue distribution of a hermitian matrix model with a Wilson loop insertion in a large representation $R$ (see Figure \ref{parametrization}) is to split the matrix $\xi$ into $g+1$ blocks $\xi_I$ of size $n_I\times n_I$ and rewrite the traces in terms of interactions among different submatrices
\ba
&&\dim R\, \langle W_R\rangle_{MM}
=\f{1}{{\cal Z}_H}\int [d \xi] \, e^{-\f{2N} \lambda \Tr \, \xi^2}
\Tr_R(e^{\xi})
\cr
&&\hskip 1cm=\f{1}{{\cal Z}_H}\int \prod_I [d\xi_I] \, e^{-\f{2N} \lambda \sum_I \Tr\,  \xi_I^2}
e^{\sum_I K_I \Tr\xi_I}\prod_{I<I'}\det \f{\left(\xi_I \otimes 1 -1\otimes 
\xi_{I'}\right)^2}{1-e^{-\xi_I}\otimes e^{\xi_{I'}}}\, ,
\label{Hermitian-R}
\ea
where $K_I$ are defined in Figure \ref{parametrization}.
At the saddle point of the integral, the eigenvalues of $\xi_{I}$
for fixed $I$ are distributed along
some interval $[e_{2I},e_{2I-1}]$.
These intervals are ordered as
\ba
e_{2g+2}<\ldots<e_1\, .
\ea
In the limit
\ba
\lambda\gg 1\,,\qquad g_{YM}^2 n_I=\Ocal(\lambda)\,,\qquad g_{YM}^2 (K_I-K_{I+1})=
\Ocal( \lambda^{1/2})\,, \label{limit}
\ea
these cuts are separated from each other by a distance of order $\sqrt\lambda$
and the exponential interactions in (\ref{Hermitian-R}) can be safely ignored.
We note that   we are studying the matrix model in the supergravity regime (\ref{limit}), where
it is meaningful to compare the matrix model computation with the corresponding computation performed 
using the dual bubbling supergravity solutions, which we carry out in the next section. Thus the matrix integral for the Wilson loop expectation value  in a large representation $R$ (see Figure \ref{parametrization}) in the supergravity regime (\ref{limit})
is given by
\ba
&& \dim R\,  \langle W_R\rangle_{MM}\cr
&&\hskip 1cm=\f{1}{{\cal Z}_H}\int \prod_I [d\xi_I] \, e^{-\f{2N} \lambda \sum_I \Tr\,  \xi_I^2}
e^{\sum_I K_I \Tr\xi_I}\prod_{I<J}\det 
\left(\xi_I \otimes 1 -1\otimes \xi_J\right)^2\, .
\label{Hermitian-R2}
\ea
This model was solved in \cite{Okuda:2008px} at large $N$.

The same trick goes through also in the normal matrix model. It is also straightforward to generalize the computation with a local  operator  insertion, which is what we need to compute (\ref{normalito}). The corresponding formula is given by
\ba
&&\dim R \, \langle W_R \Tr z^J\rangle_{MM}
\nn\\
&&\hskip 1cm=\f{1}{{\cal Z}_{N}{\cal Z}_\alpha}
\int_{[z_I,\bar z_I]=0}
\prod_I[dz_I][d\alpha_I]e^{-\f{2N}{\lambda}\sum_I\Tr(\bar z_I z_I+\alpha^2_I)}e^{\sum_I K_I\Tr\left(\f{z_I+\bar z_I}{\sqrt 2}+i\alpha_I\right)}
\nn\\
&&\hskip 3cm \times \prod_{I<I'}\f{|\mbox{det}(z_I\otimes 1-1\otimes z_{I'})|^2}{\mbox{det}\left(1-e^{-\f{z_I+\bar z_I}{\sqrt 2}-i\alpha_I}\otimes e^{\f{z_{I'}+\bar z_{I'}}{\sqrt 2}+i\alpha_{I'}}\right)}\sum_I\Tr z_I{}^J.
\label{Normal-RJ}
\ea
Similarly to the hermitian matrix model,  in the supergravity regime (\ref{limit}), the exponential
interactions in the denominator in the second line of (\ref{Normal-RJ})
can be neglected. 
Then the dynamics of $\alpha$ is decoupled and trivial as it doesn't
have the Vandermonde repulsion.
Thus we get
\ba
&&\f{\langle W_R \Tr z^J\rangle_{MM}}
{\langle W_R\rangle_{MM}}
\nn\\
&=&
\f{
\displaystyle
\int_{[z_I,\bar z_I]=0}
\prod_I[dz_I]e^{-\f{2N}{\lambda}\sum_I\Tr\bar z_I z_I}e^{\sum_I K_I\Tr\left(\f{z_I+\bar z_I}{\sqrt 2}\right)}
\prod_{I<I'}
|\mbox{det}(z_I\otimes 1-1\otimes z_{I'})|^2\sum_I\Tr z_I{}^J
}{\displaystyle
\int_{[z_I,\bar z_I]=0}
\prod_I[dz_I]e^{-\f{2N}{\lambda}\sum_I\Tr\bar z_I z_I}e^{\sum_I K_I\Tr\left(\f{z_I+\bar z_I}{\sqrt 2}\right)}
\prod_{I<I'}|\mbox{det}(z_I\otimes 1-1\otimes z_{I'})|^2}.
\nn\\
\label{Normal-RJ2}
\ea

We can study the resulting matrix model in the large $N$ saddle point approximation.
Since the submatrices of $z$ feel different
constant forces proportional to $K_I$, 
the eigenvalues of $z$ spread into
$g+1$ droplets (see Figure \ref{droplets}(c)), 
and we can easily obtain the normal matrix model saddle point equations
\be
-\f{2N}{\lambda}z_{Ii}+\f{K_I}{\sqrt 2}+\sum_{(I',i')\neq(I,i)}\f{1}{\bar z_{Ii}-\bar z_{I'i'}}=0\, , \qquad I=1,\ldots g+1, \qquad i=1,\ldots, n_I \, ,
\label{NM-SPE}
\ee
where $I$ labels 
the droplets $D_I$ and $i$ the eigenvalues inside each droplet, and $\sum_{I=1}^{g+1}n_I=N$ (see Figure \ref{parametrization}).


\subsubsection{Large \texorpdfstring{$N$}{N} solution  of the normal matrix model}

In order to solve the normal matrix model in the supergravity regime (\ref{limit}), we
 define the resolvent of the normal matrix model as
\be
\Omega(z)\equiv g^2_{YM}\sum_{I,i}\f{1}{z-z_{I,i}}
=\lambda\int_{\mathbb{C}}d^2z'\, \sigma(z',\bar z')\f{1}{z-z'}\, ,
\ee
where $\sigma(z,\bar z)$ is the eigenvalue density in the complex plane
and we use the measure $d^2 z=d({\rm Re}\, z) d({\rm Im}\,z)$.
As in any normal matrix model,
the eigenvalue density is constant in the droplets (which are then incompressible), 
as one easily sees by rewriting the large $N$ saddle point equations (\ref{NM-SPE}) as
\be
-2z+\f{g^2_{YM} K_I}{\sqrt 2}+\bar{\Omega(z)}=0
\label{saddle-Normal}
\ee
for $z\in D_I$, and using that
$\partial_z[1/(\bar z-\bar z')]=\pi \delta^2(z-z') $:
\be
\sigma(z,\bar z)=\left\{ \begin{array}{l c}  \f{2}{\pi\lambda} & 
 ~~~ \hbox{ for }  z\in D_I, \\
 & \\
0 &  ~~~ \hbox{ for }  z\notin D_I. 
\end{array} \right.
\ee
The resolvent then becomes 
\be
\Omega(z)=\f{2}{\pi}\int_{D=\bigcup_{I}D_I}d^2z'\, \f{1}{z-z'}\, .
 \ee

Our task is to find a function $\Omega(z)$ and $g+1$
simply connected regions $D_I$ such that
$\Omega(z)$ is holomorphic outside $D=\cup_I D_I$, equation 
(\ref{saddle-Normal}) is satisfied on the boundaries $\p D_I$,
and 
\ba
\Omega(z)= \f{\lambda}z+\Ocal(z^{-2})~~~~~ \hbox{ as } z\ra\infty\, .
\ea
As we saw in equation (\ref{finalmani}), the normal matrix model is simply a rewriting
of the hermitian matrix model if we ignore the $\Tr z^J$ insertion, to which
the eigenvalues do not back-react anyway.
So we should expect that the large $N$ solutions
of the hermitian and the normal matrix models are related.

The hermitian matrix model
(\ref{Hermitian-R2})
has been solved in   \cite{Okuda:2008px}. 
The resolvent
\ba
\omega_1(\zeta)=g_{YM}^2 \sum_I  \oo{\zeta-\xi^{(I)}} 
\label{omega1-def}
\ea
is given as the indefinite integral
\ba
\omega_1(\zeta)=\int^\zeta_\infty
\left(2-2\f{a_{g+1}(\zeta')}{\sqrt{H_{2g+2}(\zeta')}}\right)d\zeta' .
\label{omega1}
\ea
Here $a_{g+1}(\zeta)$ and 
\ba
H_{2g+2}(\zeta)=\prod_{i=1}^{2g+2}(\zeta-e_i)
\ea
 are monic polynomials
of degree $g+1$ and $2g+2$, respectively.
Their coefficients are
determined by the constraints described in \cite{Okuda:2008px}.
These constraints guarantee that
the integration contour in (\ref{omega1}) is arbitrary
as long as it does not cross any of the cuts $[e_{2I},e_{2I-1}]$.
The integrand is a meromorphic one-form on the hyperelliptic curve
given by the equation
\ba
w^2=H_{2g+2}(\zeta). \label{hyperel}
\ea

\begin{figure}[tb]
\begin{center}
\begin{tabular}{ccc}
\psfrag{zeta}{$\zeta$}
\includegraphics[scale=.35]{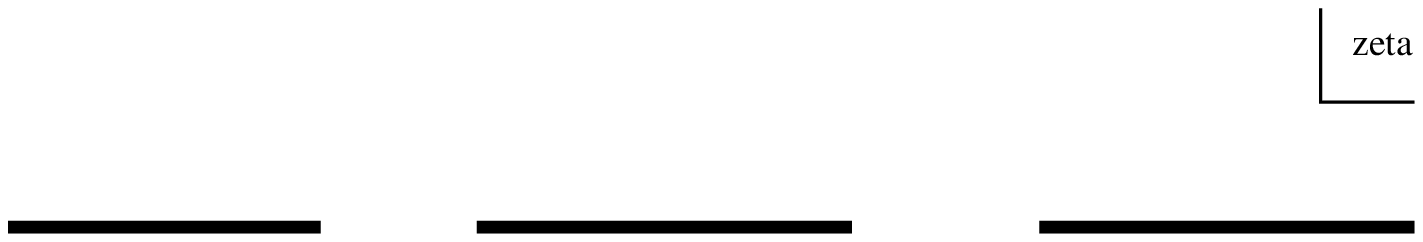}
\vspace{-8mm}
&
&
\psfrag{x}{$\xi$}
\psfrag{rho}{$\rho(\xi)$}
\includegraphics[scale=.35]{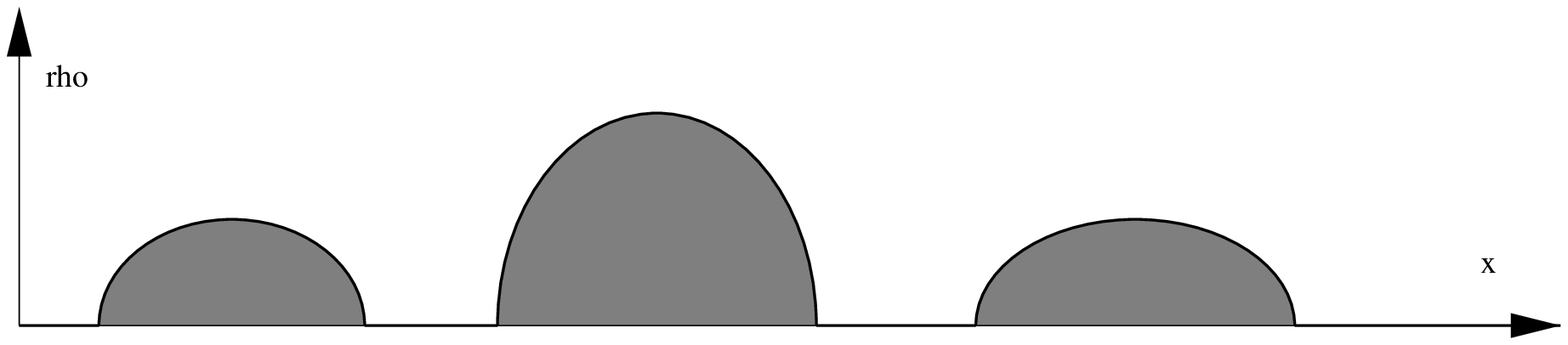} 
\\
(a)&&(b)
\\
&&
\end{tabular}
\begin{tabular}{c}
\psfrag{z}{$z$}
\includegraphics[scale=.35]{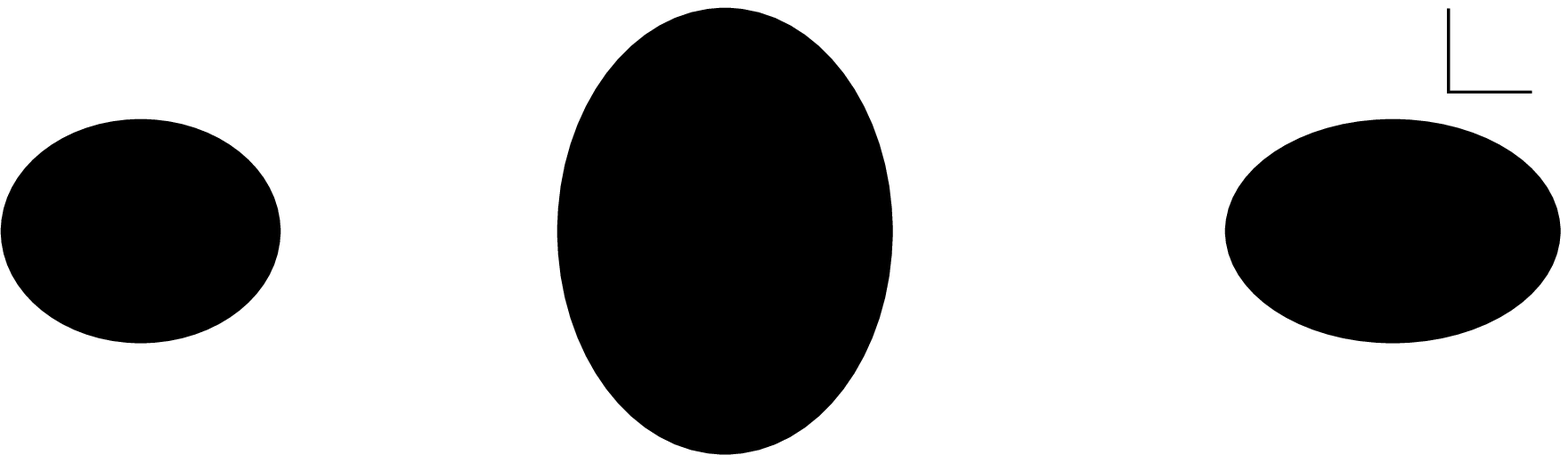} 
\\
(c)
\end{tabular}
\caption{
(a) The branch cuts $[e_{2I}, e_{2I-1}]$ ($I=1,\ldots, g+1$) on 
the $\zeta$ plane $\C$
for the the hyperelliptic curve (\ref{hyperel}) in the $g=2$ case.
(b) The eigenvalue density $\rho(\xi)$ of 
the hermitian matrix model (\ref{Hermitian-R}).
(c) The corresponding 
eigenvalue distribution (droplets) of the normal matrix model (\ref{Normal-RJ}).
The shape of a droplet is given by $\rho(\xi)$.
}
\label{droplets}
\end{center}
\end{figure}

Let us define a holomorphic function $\omega_2(\zeta)$
by analytically continuing the resolvent $\omega_1(\zeta)$
to the second sheet
along the $(g+1)$-th branch cut:
\ba
\omega_2(\xi\pm i\epsilon)=\omega_1 (\xi\mp i\epsilon)~~~~~\hbox{ for }
\xi \in [e_{2g+2},e_{2g+1}]\, .
\ea
In other words,
\ba
\omega_2(\zeta)=\omega_1(e_{2g+2})+
\int_{e_{2g+2}}^\zeta\left(
2+\f{a_{g+1}(\zeta')}{\sqrt{H_{2g+2}(\zeta')}}
\right) d\zeta', \label{omega2}
\ea
where again the contour should not cross any cut.
It then follows that
\be
\omega_{2}(\xi\pm i\epsilon)
-\omega_{1}(\xi\mp i\epsilon)=\oint_{B_I} d\zeta 
\left(2-2\f{a_{g+1}(\zeta)}{\sqrt{H_{2g+2}(\zeta)}}\right)=g^2_{YM} K_I,
\label{omega12}
\ee
where the contour $B_I$ goes around the interval $[e_{2g+1},e_{2I}]$
, from $e_{2I}$ to $e_{2g+1}$ on the first sheet and from $e_{2g+1}$ to
$e_{2I}$ on the second sheet.
The last equality in (\ref{omega12}) is one of the constraints
\cite{Okuda:2008px}.

We claim that the solution to our complex analysis problem determining the resolvent of the normal matrix model
is given by
$\Omega(z)=(1/\sqrt 2) \omega_1\circ\omega^\mo_2(2\sqrt 2 z)$,
or 
\ba
z&=&\oo{2\sqrt 2}\omega_2(\zeta), \label{z-omega2}
\\
\Omega(z)&=&\oo{\sqrt 2} \omega_1(\zeta).\label{z-omega1}
\ea
First, one can see from (\ref{omega2}) that (\ref{z-omega2}) maps 
$\C-\cup_I [e_{2I},e_{2I-1}]$ to the complement
of some droplets, identified with $D_I$.
That is, it maps Figure \ref{droplets}(a) to Figure \ref{droplets}(c).
For $z\in \p  D_I$ with ${\rm Im}\, z \lessgtr 0$,
then
\ba
-2z+\f{g_{YM}^2 K_I}{\sqrt 2}
+\bar{\Omega(z)}
=
-\oo{\sqrt{2}} \omega_2(\xi\pm i\epsilon)+\f{g_{YM}^2 K_I}{\sqrt 2}
+\oo{\sqrt 2}\bar{\omega_1
(\xi\pm i\epsilon)}
=0,
\ea
where we used that $\bar{\omega_1
(\xi\pm i\epsilon)}=\omega_1(\xi\mp i\epsilon)$.
Thus (\ref{saddle-Normal}) is indeed satisfied!

In summary, we have found a normal matrix model eigenvalue distribution in terms of the hermitian matrix model. 
This  now allows to calculate $\Xi_{R,J}$ in (\ref{Xi}) in terms of the moments of the hermitian matrix model 
eigenvalue distribution. We will then compare this with the supergravity computation in the next section, where exact agreement will be found.


\subsubsection{Correlators as moments 
in the normal matrix model}

All the information about the representation $R$ of the Wilson loop is  encoded in the moments of the matrix model eigenvalue distribution, which for the hermitian matrix model are defined by
\be
\langle \xi^n\rangle \equiv \rho_n \equiv \int d\xi \, \rho(\xi)\, \xi^n\, .
\label{herm-mom}
\ee
Here  $\rho(\xi)$ is the hermitian matrix model eigenvalue distribution. 
We want to express the normal matrix model moments
\ba
\langle z^J\rangle=\int d^2 z\ \sigma(z,\bar z) z^J
\ea
 in terms of the $\langle \xi^n\rangle$'s.

The hermitian matrix model resolvent $\omega_1(\zeta)$ can be expanded in moments of the eigenvalue distribution
\be
\omega_1(\zeta)=\lambda\int d\xi\, \rho(\xi)\f{1}{\zeta-\xi}
=\lambda\sum_{n=0}^\infty\f{\langle \xi^{n}\rangle}{\zeta^{n+1}}\, ,
\label{expande}
\ee
with $\langle \xi\rangle=0$.\footnote{
\label{COM}
This imposes a center of mass condition on the distribution. So far we have focused on the $U(N)$ case. For $SU(N)$ 
  gauge group, we
need to restrict the integrated hermitian and complex matrices.
Since the $U(1)$ part is decoupled,
in the large $N$ limit,
the net effect is to shift the eigenvalue distributions 
so that the
average eigenvalues vanish: $\langle \xi\rangle=\langle z\rangle=0$.} This constraint can be interpreted as arising from considering ${\cal N}=4$ super Yang-Mills with $SU(N)$ gauge group.

From  the $SU(N)$ saddle point equation on the $(g+1)$-th cut, 
we have for any $\zeta$ that \cite{Okuda:2008px}
\be
\omega_1(\zeta)+\omega_2(\zeta)=4\zeta+g_{YM}^2 |R|/N\,,
\ee
where $|R|$ is the total number of boxes in the Young tableau corresponding to the 
representation $R$ (see Figure \ref{parametrization}).
After shifting $z$ by $\sqrt 2g_{YM}^2 |R|/4N$ in (\ref{z-omega2}),
we get that
\be
z=\sqrt 2 \zeta-\f{1}{2\sqrt 2}\omega_1(\zeta)\,.
\label{invers}
\ee
This relation can be inverted recursively using (\ref{expande}, \ref{invers}) to obtain
\ba
\zeta=
&&
\frac{z}{\sqrt{2}}+\frac{\lambda }{2 \sqrt{2} z}
+\frac{4\lambda  \langle \xi^2\rangle-\lambda^2}{4\sqrt 2 z^3}
+\frac{\lambda  \langle \xi^3\rangle}{z^4}
\nn\\
&&
~~~~~+\frac{\sqrt{2}
   \lambda ^3-8 \sqrt{2} 
\langle \xi^2\rangle \lambda ^2+8 \sqrt{2} \langle \xi^4\rangle\lambda }{8 z^5}
+\Ocal(z^{-6}).
\label{relacio}
\ea
We can now write the resolvent of the normal matrix model in terms of the hermitian matrix
model moments of the eigenvalue distribution by combining (\ref{z-omega1}, \ref{relacio})
\ba
\Omega(z)
&=&
\lambda \sum_{J=0}^\infty \f{\langle z^J\rangle}{z^{J+1}}
\nn\\
&=&\f{\lambda}{z}+\f{2\lambda\langle \xi^2\rangle-\f{1}{2}\lambda^2}{z^3}+\f{2\sqrt 2\lambda\langle\xi^3\rangle}{z^4}+\f{4\lambda\langle\xi^4\rangle-4\lambda^2\langle\xi^2\rangle+\f{1}{2}\lambda^3}{z^5}
 \cr
 &&
~~+\f{4 \sqrt{2} \lambda  \langle \xi^5\rangle-5 \sqrt{2} \lambda ^2 
\langle \xi^3\rangle}{z^6}
\cr &&
~~+\f{8 \langle \xi^6\rangle \lambda-12 \langle \xi^4\rangle \lambda ^2-6 
\langle \xi^2\rangle^2 \lambda^2+\frac{15}{2} \langle \xi^2\rangle \lambda ^3-\frac{5}{8} \lambda ^4 }{z^7}
+{\cal O}\left(z^{-8}\right).
\label{Omega-expansion}
\ea
This allows us to express the normal matrix model moments 
$\langle z^J\rangle$ in terms of the moments of the hermitian matrix
 model $\langle\xi^n\rangle\equiv \rho_n$.  Notice that non-linearities in the moments start to appear only at order ${\cal O}(z^{-7})$.
The first few moments of the normal matrix model are given in terms of the moments of the hermitian matrix model by
\ba
\langle z^2 \rangle& = & 2 \langle \xi^2\rangle -\f{\lambda}{2} 
= 2\Delta \rho_2\, , \cr
\langle z^3 \rangle& = & 2\sqrt 2  \langle \xi^3\rangle 
= 2\sqrt 2\Delta \rho_3\, , \cr
\langle z^4 \rangle& = & 4 \langle \xi^4\rangle -4\lambda \langle \xi^2\rangle +\f{\lambda^2}{2} 
= 4\left(\Delta \rho_4-\lambda \Delta \rho_2\right)\,,
\label{moments-final}
\ea
where $\Delta \rho_n\equiv \rho_n-\rho^0_n$.
Here we have used that the moments $\rho^0_n$ of the Wigner semi-circle distribution law, which has the eigenvalue distribution $\rho^{0}(\xi)=
(2/ \pi \lambda)\sqrt{\lambda- \xi^2}$, are given by  $\rho_2^0=\lambda/4$, 
$\rho_3^0=0$ and $\rho_4^0=\lambda^2/8$. As we shall see in the next section, the fact that the correlators are given 
by moments relative to the Wigner semi-circle distribution has a corresponding statement in supergravity, where 
correlators are captured by devitations away from the $AdS_5\times S^5$ vacuum solution, which in the parametrization of \cite{D'Hoker:2007fq} corresponds to the Wigner semi-circle distribution law.

We are now ready to write the result of our computation of the correlator coefficients $\Xi_{R,J}$ 
in the supergravity regime (\ref{limit}). From (\ref{Xi}), we find that
\ba
\Xi_{R,2}&=&\sqrt 2 \f{N}\lambda \Delta \rho_2\, ,\nn\\
\Xi_{R,3}&=& 2\sqrt{\f{2}3}\f{N}{\lambda^{3/2}} \Delta \rho_3\, ,\nn\\
\Xi_{R,4}&=& 2\f{N}{\lambda^2} (\Delta \rho_4 -\lambda \Delta\rho_2)\, .
\label{Xi234}
\ea
The expression (\ref{Xi234}), together with (\ref{WO-AdS}), is the final result of the gauge theory computation    for the correlators between a half-BPS Wilson loop  and the chiral primary operators in ${\cal N}=4$ SYM  and represents a prediction for supergravity, as we have solved the matrix model in the supergravity regime (\ref{limit}).

Moreover, using our derivation of the relation between the correlator of the Wilson loop with the stress tensor and the correlator with the dimension two chiral primary operator (\ref{relacioentre}) we also obtain 
the correlator with the
stress tensor in terms of the hermitean matrix model.
Namely, we have that 
\ba
h_W&=& 
-\f{N}{3\sqrt 2 \pi^2} \Xi_{R,2}=-\f{N^2}{3\pi^2 \lambda}\Delta \rho_2\, .
\label{hW2}
\ea
and the correlator is given by (\ref{T-ads}).

In Section \ref{sugra-sec}, we will calculate the same correlation functions 
using the dual bubbling supergravity solutions and show that we get precise agreement.


\subsection{Correlator of Wilson loop  with the stress tensor from S-duality}
\label{stress-sec}

We have already shown 
in subsection 2.3
an exact  relation between the correlator  of the Wilson loop with the stress tensor and the correlator of the Wilson loop with dimension two chiral primary operators. Using this result, we have computed the strong coupling result of this correlator in terms of the hermitian matrix model second moment of the eigenvalue distribution (\ref{hW2}).

Here we calculate the stress tensor correlator at strong coupling in an alternative way. 
This involves considering the correlator of the stress tensor with a 't Hooft line $T_R^{\rm line}$ in the semiclassical gauge  theory and then S-dualizing. This turns the 't Hooft line into a Wilson line and exchanges the weak coupling regime with the strong coupling regime. So we need to calculate the semiclassical scaling weight  $h_T$ for the 't Hooft line $T_R^{\rm line}$, which captures the correlator of a 't Hooft line operator with the stress tensor as in (\ref{T-line}) (see also \cite{Kapustin:2005py}). We find exact agreement, providing a non-trivial test of S-duality in ${\cal N}=4$ super Yang-Mills. 

The bosonic action of ${\cal N}=4$ super Yang-Mills is given by
\ba
S=\oo{g_{YM}^2}\int d^4 x\sqrt g\, \Tr\left(
\half F_{\mu\nu}F^{\mu\nu} +D_\mu\phi^i D^\mu\phi^i 
+\f{R}{6}\phi^i\phi^i-\frac{1}{2}[\phi^i,\phi^j]^2
\right).
\ea
The bosonic contribution to the stress tensor of the theory is given by 
\ba
T_{\mu\nu}&=& \f{2}{\sqrt g}\f{\delta S}{\delta g^{\mu\nu}}
\nn\\
&=&
\f{2}{g_{YM}^2} \Tr\left(
F_{\mu\rho}F_{\nu}^{~\rho}-\oo 4 g_{\mu\nu}F_{\rho\sigma} F^{\rho\sigma}
\right)+
\f{2}{g_{YM}^2}\Tr\left(
D_\mu \phi^i D_\nu \phi^i
-\half g_{\mu\nu} D_\rho\phi^i D^\rho\phi^i\right.
\cr &&\left. \hskip 1cm-\f{R}{12} g_{\mu\nu}\phi^i \phi^i+\oo 6 R_{\mu\nu}\phi^i\phi^i+\oo 6 (g_{\mu\nu} D^2-D_\mu D_\nu)(\phi^i\phi^i)+\frac{1}{4} g_{\mu\nu}
 [\phi^i,\phi^j]^2
\right)
.\cr 
\label{T-S}&&
\ea

We want to compute the correlator $\langle T_R^{\rm line} T_{\mu\nu}(x)\rangle$. In the semiclassical approximation, this is found by evaluating the stress tensor  (\ref{T-S})  on the gauge field configuration produced by the insertion of $T_R^{\rm line}$, which is given by (see {\it e.g.} \cite{Kapustin:2005py})
\ba
F=\half B\, {\rm vol}_{S^2}\, ,\qquad \phi= \f{B}{2l}\, .
\ea
Here ${\rm vol}_{S^2}$ is the volume form on the $S^2$  surrounding the 't Hooft line, $l$ is the distance from the 't Hooft line and $B$ is the highest weight vector for the representation $R$ (see Figure \ref{parametrization}).

The semiclassical scaling weight for the 't Hooft line operator $T_R^{\rm line}$  is given by (c.f. (\ref{T-line})) 
  \ba
h_T=-\oo{3 g_{YM}^2  }\Tr (B^2)   + {\rm corrections}
\label{T-S1}
\, ,
\ea
where the ${\rm corrections}$ are due to loop effects, that would be interesting to compute.
For  gauge group $SU(N)$, $B$ -- the highest weight vector of the representation $R$ -- is given by
\ba
B={\rm diag}\left(R_1-|R|/N, R_2-|R|/N, \ldots, R_N-|R|/N\right),
\ea
so that  (see Figure 1)
\ba
\Tr B^2=\sum_I n_I (K_I-|R|/N)^2.
\ea
Therefore, we have found that in the semiclassical approximation the scaling weight of the 't Hooft line operator $T_R^{\rm line}$ in a representation $R$ is given by
\ba
h_T=-\oo{3 g_{YM}^2}\sum_I n_I (K_I-|R|/N)^2.
\label{vivatoo}
\ea

We are interested in computing the scaling weight  of the corresponding Wilson line $W_R^{\rm line}$  $h_W$ at strong coupling. S-duality is expected to exchange the  't Hooft line $T_R^{\rm line}$ with the Wilson line $W_R^{\rm line}$, as well as exchange the weak coupling regime with the strong coupling regime.
Therefore, for the Wilson loop scaling weight $h_W$ at strong coupling, we should S-dualize the 't Hooft loop result (\ref{vivatoo}) by replacing $g_{YM}^2\ra 16\pi^2/g_{YM}^2$.
The S-dual scaling weight of the 't Hooft line -- which we denote by $h^S_T$ -- is then given by
\ba
h^S_T=-\f{g_{YM}^2}{48\pi^2 }\sum_I n_I(K_I-|R|/N)^2+{\rm corrections}\,.
\label{tooft}
\ea

On the other hand, the strong coupling result we obtained using the normal matrix model for the scaling weight of the Wilson loop $W_R^{\rm line}$ is (\ref{hW2})
\ba
h_W&=& -\f{N^2}{3\pi^2 \lambda}\Delta \rho_2\, .
\label{hW22}
\ea
We now note that
\ba
 \langle z^2\rangle=2 \Delta \rho_2 =\f{g_{YM}^4}{8 N}\sum_I n_I 
(K_I-|R|/N)^2
\label{limiraro}
\ea
in the limit that the cuts in the eigenvalue distribution of the matrix model are widely separated,
so that 
\ba\rho(\xi)=\sum_I \f{n_I}N  \delta\left(\xi-\f{g_{YM}^2 (K_I-|R|/N)}4\right).
\ea
 In this particular limit, the expression for the S-dual of the 't Hooft loop scaling weight (\ref{tooft}) 
agrees precisely with the computation of the scaling weight of the Wilson loop in the strong coupling regime, obtained by combining (\ref{hW22}) and (\ref{limiraro}).
This is a non-trivial quantitative test of S-duality for $\Ncal=4$ $SU(N)$
super Yang-Mills.

It is desirable to understand why  quantum corrections
are suppressed in this limit and to explicitly compute
them. Once these are included in the 't Hooft loop computation, the agreement we found could be 
extended.
We hope to come back to these issues in the near future.


\section{Correlators from supergravity}
\label{sugra-sec}

In this section we compute the correlation functions of a half-BPS circular Wilson loop with the chiral primary operator  ${\cal O}_J$ for $J=2,3,4$ and with the stress tensor of ${\cal N}=4$ super Yang-Mills using the  bubbling supergravity backgrounds 
found in \cite{Yamaguchi:2006te,Lunin:2006xr,D'Hoker:2007fq}. These geometries are regular solutions of ten-dimensional type IIB supergravity that are asymptotically $AdS_5\times S^5$  and provide the gravitational description of all half-BPS circular Wilson loops in ${\cal N}=4$ super Yang-Mills. They capture the complete backreaction of the configuration of D5 or D3 branes in $AdS_5\times S^5$   describing a half-BPS Wilson loop  in an arbitrary representation $R$ of the gauge group \cite{Gomis:2006sb} (see also \cite{Drukker:2005kx, Yamaguchi:2006tq, Gomis:2006im}).

For such ten-dimensional asymptotically  $AdS_5\times S^5$ solutions  there is a well-defined procedure, developed in \cite{Skenderis:2006uy,Skenderis:2007yb},\footnote{For some of the previous work on holographic renormalization see \cite{deHaro:2000xn,Skenderis:2000in,Bianchi:2001de,Bianchi:2001kw,Papadimitriou:2004rz}. A nice review of these topics can be found in \cite{Skenderis:2002wp}.} to extract the one-point functions of local operators in the state produced by the  operator that the bubbling solution describes. Using this method, we will be able to obtain the correlators  of the Wilson loop from the asymptotic expansion of various bulk fields, which we compute using    the bubbling supergravity solution. We will find exact agreement with the 
strong coupling computation in gauge theory performed in Sections 
\ref{sec-sym} and \ref{CFT-sec} using matrix models and S-duality.


\subsection{Review of the bubbling solution}

We start by briefly reviewing the bubbling solution found in  \cite{Yamaguchi:2006te,Lunin:2006xr,D'Hoker:2007fq}, using the elegant parametrization of the solution found in \cite{D'Hoker:2007fq} (to which we refer the reader for more details).  As is well-known, a half-BPS circular Wilson loop preserves an $Osp(4^*|4)$ subalgebra of the $PSU(2,2|4)$ algebra  of  symmetries of ${\cal N}=4$ super Yang-Mills. The 
$SO(2,1) \times SO(3) \times SO(5)$ bosonic symmetries in $Osp(4^*|4)$  are realized in the ten-dimensional supergravity solution by writing the ten-dimensional metric  as  an $AdS_2 \times S^2 \times S^4$ fibration over a two-dimensional base manifold and by writing the most general ansatz for the other supergravity fields compatible with this symmetry. 

The metric describing a half-BPS Wilson loop is then given by
\be
ds^2 = f_1 ^2 ds^2 _{AdS_2} + f_2 ^2 ds^2 _{S^2} + f_4^2 ds^2 _{S^4}
+ 4 \rho^2 (dx^2+dy^2)
\label{metric0}
\ee
where the warp factors $f_1,\, f_2,\, f_4$, and $\rho$  are real functions on the base.\footnote{
The same symbol $\rho$
denotes both the eigenvalue density as well as a component
of the metric. 
The distinction should be clear from the context.
} The warp factors and the supergravity fluxes can be completely expressed in terms of two harmonic functions $h_1$ and $h_2$ on the base \cite{D'Hoker:2007fq}.
In \cite{Okuda:2008px}, a precise relation has been found between these harmonic functions and the data that control the spectral curve of the hermitian matrix model, which  captures the vacuum expectation value of a half-BPS Wilson loop in ${\cal N}=4$ super Yang-Mills \cite{Erickson:2000af,Drukker:2000rr,Pestun:2007rz}. The mapping is given by \cite{Okuda:2008px}
\ba
h_1=\f{i\alpha'}{8g_s}\left(2(z-\bar{z})-(\omega_1-\bar{\omega}_1) \right)\, , \qquad h_2=\f{i\alpha'}4 (z-\bar{z})
\label{h1-h2}
\ea
where
\ba
 z=-i\sqrt\lambda \sinh (x+iy)\,,\qquad
\label{mapeo}
\ea
and $\omega_1$ is the hermitian matrix model resolvent introduced   in equation
(\ref{omega1-def}), and $z$ is the spectral parameter of the resolvent. Notice that we use from now on, for graphical clarity, the letters $z$, $x$, and $y$ for the hermitian matrix model variables (whereas in the previous section we have used Greek letters for the hermitian matrix model and Latin letters for the normal matrix model). The hermitian matrix model resolvent is given by 
\ba
\omega_1=g^2_{YM}\Tr{1\over z-M}\equiv \lambda \int {\rho(x)\over z-x}.
\ea
We note that the information of the representation of the Wilson loop is encoded in the resolvent $\omega_1$, which depends non-trivially on the choice of representation $R$ of the Wilson loop, while the harmonic function $h_2$ is universal and independent of the representation.

It is convenient to define the following combinations of the harmonic functions $h_1$ and $h_2$ and their derivatives
\ba
 V&\equiv&{1\over 2}
{\partial h_1\over \partial y}{\partial h_2\over \partial x}
-{1\over 2}{\partial h_1\over \partial x}{\partial h_2\over \partial y}\,, \qquad
\cr
W&\equiv&{1\over2}
{\partial h_{1}\over \partial x}{\partial h_{2}\over \partial x}
+{1\over2}
{\partial h_{1}\over \partial y}{\partial h_{2}\over \partial y}\, ,\cr
 N_1&\equiv&{1\over2}h_1 h_2
\left(\left({\partial h_{1}\over \partial x}\right)^2+
\left({\partial h_{1}\over \partial y}\right)^2
\right)-h^2_1W \, ,\cr
N_2&\equiv&{1\over2}h_1 h_2
\left(\left({\partial h_{2}\over \partial x}\right)^2+
\left({\partial h_{2}\over \partial y}\right)^2
\right)-h^2_2W 
\label{nicefuncts}
\, .\ea
The warp factors in the metric (\ref{metric0}) are  then given by \cite{D'Hoker:2007fq}
\ba
f^2_1&=&\left(-4\sqrt{-{N_2\over N_1}}h^4_1{W\over N_1}\right)^{1/2}\, ,\cr
f^2_2&=&\left(4\sqrt{-{N_1\over N_2}}h^4_2{W\over N_2}\right)^{1/2}\, ,\cr
f^2_4&=&\left(4\sqrt{-{N_1\over N_2}}{N_2\over W}\right)^{1/2}\, ,\cr
\rho^2&=&\left(-{W^2 N_1N_2\over h_1^4h_2^4}\right)^{1/4}
.\ea

The RR four-form  can be read off from
\ba
dC_{(4)}&=&
-dj_{1}\hat{e}^{0123}+dj_{2}\hat{e}^{4567}\cr
&\equiv&-dj_{1}\hat{e}^{0123}+(F_xdx+F_ydy)\hat{e}^{4567}
\label{fourrrr}
,\ea
where $j_2$, as shown in  \cite{Okuda:2008px},  is given by
\ba
dj_2&=&-if_4^4 \rho(f_zdw-f_{\bar{z}}d\bar{w})\cr
&=&\left[
{\partial \over \partial x}
\left(h_1h_2 {V\over W}\right)
+3\left(
h_1{\partial h_2\over \partial y}
-h_2{\partial h_1\over \partial y}
\right)\right]dx\cr
&&+
\left[
{\partial \over \partial y}
\left(h_1h_2 {V\over W}\right)
-3\left(
h_1{\partial h_2\over \partial x}
-h_2{\partial h_1\over \partial x}
\right)\right]dy
\label{5form}
.\ea
In the expression for the four-form gauge field (\ref{fourrrr}), $\hat{e}^{0123}$ is the $AdS_2\times S^2$ volume form and $\hat{e}^{4567}$ is the $S^4$ volume form, both with unit radius.

The complete bubbling supergravity solution also excites the dilaton and the RR and NS-NS two-form gauge fields of type IIB supergravity. However, we will not need their explicit expressions in this paper, as we will later show that these fluxes do not contribute to the correlation functions we compute. Their explicit expressions can be found in \cite{D'Hoker:2007fq}.

In order to calculate the correlation function of a half-BPS Wilson loop and a local operator, we must study the deviations of the bubbling supergravity solution from the $AdS_5\times S^5$ vacuum solution. Our task will then be to extract  the various correlation functions from the deviations from this vacuum. Therefore, we first consider the eigenvalue distribution corresponding to $AdS_5\times S^5$. In the language of the hermitian matrix model, the eigenvalue distribution for this case is the Wigner semi-circle law
\ba
\omega_1^{(0)}=2z-2\sqrt{z^2-\lambda}\, .
\ea
From equations (\ref{h1-h2}, \ref{mapeo}) we get that the harmonic functions corresponding to $AdS_5\times S^5$  are given by
\ba
h_{1}^{(0)}=4c^2\cosh{x}\cos{y}\,,\qquad
h_{2}^{(0)}=4c^2\sinh{x}\cos{y},
\ea
where $c^2=\sqrt{\lambda}\alpha'/8$ (we have set the background dilaton to zero for simplicity).
For $AdS_5\times S^5$ the functions defined in (\ref{nicefuncts}) are given by
\ba
V^{(0)}&=&-4c^4\sin{2y}\, ,\cr
W^{(0)}&=&4c^4\sinh{2x}\, , \cr
N_{1}^{(0)}&=&-64c^8\cos^4{y}\sinh{2x}\, , \cr
N_{2}^{(0)}&=&64c^8\cos^4{y}\sinh{2x}\, ,
\ea
while the $AdS_5 \times S^5$ warp factors are\footnote{Here $y\in[-\pi/2,\pi/2]$.}
\ba
f^{(0)}_{1}=2\sqrt 2c\cosh{x}\,,\qquad
f^{(0)}_{2}=2\sqrt 2c\sinh{x}\,,\qquad
f^{(0)}_{4}=2\sqrt 2c\cos{y}\,,\qquad
\rho^{(0)}=\sqrt 2c\,. \cr
&& 
\ea
These warp factors give rise to the  $AdS_5 \times S^5$ metric 
\ba
ds^2=8c^2\left(\cosh^2{x}\, ds^2_{AdS_2}+dx^2+\sinh^2{x}\, ds^2_{S^2}\right)
+8c^2\left(dy^2+\cos^2{y}\, ds^2_{S^4}\right),
\label{vaccc}
\ea
where $AdS_5$ is foliated by $AdS_2 \times S^2$ slices while the $S^5$ is foliated by $S^4$'s, a slicing that makes manifest the symmetries of the half-BPS Wilson loop. For the vacuum solution only the RR four-form is excited and
\ba
dj_{2}^{(0)}=-64c^4\cos^4{y}\, dy.
\ea


\subsection{Kaluza-Klein holography}

We recall now how to  holographically compute the one-point functions of local operators from asymptotically  $AdS_5\times S^5$ supergravity geometries.\footnote{The procedure is actually more general and applies to any asymptotically $AdS_p\times X^q$ solution, with $X^q$ being a compact manifold.} For more details we recommend  \cite{Skenderis:2006uy,Skenderis:2007yb}.

Given an asymptotically  $AdS_5\times S^5$ supergravity solution, one needs to expand all the ten-dimensional fields excited in the solution  in a complete basis of spherical harmonics on the $S^5$. This produces in general an infinite number of five-dimensional fluctuation fields. These fluctuation modes are, however, not independent. Some of them are in fact related to each other by the action of ten-dimensional diffeomorphisms, which give rise to non-linear gauge transformations on the five-dimensional fluctuations. Instead of gauge fixing these symmetries  (by going for example to de Donder gauge) as it is usually done in the study of the spectrum, it is more convenient to construct gauge invariant combinations of the fluctuations. This is because  generic ten-dimensional  supergravity solutions, such as the bubbling Wilson loop backgrounds we are considering, are generally not in de Donder gauge. 
The equations of motion solved by the gauge invariant fluctuations are nevertheless   the same   as those of the fluctuations in the de Donder gauge.

The gauge invariant combinations of fluctuations  obey in general non-linear equations of motion containing higher derivative terms, just like the fluctuations in the de Donder gauge do. These equations of motion with the higher derivatives, however, cannot be obtained from a local five-dimensional action. In order to perform holographic computations of correlators using supergravity, we want to rewrite the bulk action using a local bulk action. 
This  can be accomplished  by performing a   {\it Kaluza-Klein reduction map}, a non-linear map between solutions to the ten-dimensional equations of motion, $\psi_{10d}$, and solutions to the five-dimensional ones, $\Psi_{5d}$, which can be schematically expressed as
\be
\Psi_{5d} = \psi_{10d} + {\cal K}\,\psi_{10d}\,\psi_{10d} + \ldots
\label{KKmap}
\ee 
where ${\cal K}$ is some operator containing also derivatives and the ellipses denote higher order combinations of the 10-dimensional fields and their derivatives. Notice that, in principle, all Kaluza-Klein modes are kept in the reduction map. However, in practice, when computing the expectation value of some operator with a given dimension, only a finite number of modes will contribute, giving an effective truncation of the Kaluza-Klein tower. This will also limit the number of non-linear terms in (\ref{KKmap}) that one needs to compute. In general, the higher the dimension of the operator one considers, the more terms have to be turned on. For example, we shall see that for the dimension four chiral primary operators only two terms in the map are needed: a linear term with dimension four and a  term quadratic in the fields dual to dimension two operators.

At this point, the new equations in the reduced fields $\Psi_{5d}$ can be integrated into a local five-dimensional action and one can use the general holographic rules to compute gauge theory correlation functions from supergravity \cite{Maldacena:1997re,Gubser:1998bc,Witten:1998qj}.
The local five-dimensonal action does  suffer from infrared divergences and has  to be regularized by the addition of appropriate boundary counterterms and then evaluated on-shell. Differentiating the regularized on-shell action with respect to appropriate sources in the spirit of \cite{Maldacena:1997re,Gubser:1998bc,Witten:1998qj}  yields the renormalized one-point functions of the dual gauge theory operators. These are  related  to certain coefficients in the near boundary expansion of the bulk fields, which correspond to the normalizable fluctuation mode of the field. The higher the dimension of the operator, the deeper in $AdS$ space one needs to dig to extract its one-point function.

In this paper we are interested in computing the one-point functions of ${\cal O}_J$ for $J=2,3,4$ and the stress tensor  in the state created by a half-BPS Wilson loop. The calculation of these relatively low dimension operators enjoys a great simplification. In order to calculate their one-point functions we can neglect all the supergravity fields excited by the bubbling solution except for the metric and the RR four-form gauge field. The fluctuations that arise from the other fields (the dilaton and the RR and NS-NS two-form gauge fields) do not enter in the calculation of the one-point function of the local operators under study. Technically, the reason this occurs is that the fluctuations coming from the dilaton, the RR and NS-NS two-form gauge field fall off too fast near the $AdS_5\times S^5$ boundary and therefore do not enter into the  Kaluza-Klein reduction map
(\ref{KKmap}) for the dual fluctuations. We can therefore use the formulas relating five-dimensional fluctuations and one-point functions of dual gauge theory operators derived in  \cite{Skenderis:2006uy,Skenderis:2007yb}.

Without further ado, we setup our computation by expanding the bubbling Wilson loop solution in fluctuations around the $AdS_5\times S^5$ background. We define the fluctuations in the metric (\ref{metric0}) as 
\ba
ds^2&=&(f^{(0)}_{1})^2(1+\Delta_1)\,ds^2_{AdS_2}+
(f^{(0)}_{2})^2(1+\Delta_2)\,ds^2_{S^2}
\cr &&+(f^{(0)}_{4})^2(1+\Delta_4)\,ds^2_{S^4}
+4(\rho^{(0)})^2(1+\Delta_{\rho})\,(dx^2+dy^2).
\label{met-fluc}
\ea
Physical quantities, such as the correlation functions we are after, are encoded in the asymptotic expansion of these functions for large $x$.

Since all the information of the matrix model is given by the resolvent, the fluctuations
$\Delta_1,\, \Delta_2,\, \Delta_4$, and $\Delta_{\rho}$  should  depend only on the matrix model resolvent $\omega_1$.
So we first need to determine the relation between 
the resolvent and the  asymptotic form of the $\Delta$'s. We start by expanding $h_1$ for large $x$
\ba
h_{1}&=&4c^2\cosh{x}\cos{y}\left(1+\sum^\infty_{m=3}c_{m}(y)e^{-mx} \right)
\label{grahone}
\, ,\ea
whereas $h_2$, being independent of $\omega_1$, remains equal to its background value $h_{2}^{(0)}$ corresponding to $AdS_5\times S^5$.
The corrections start from order $e^{-3x}$ because the background ($\cosh x$)
includes $e^{x}$ and $e^{-x}$ and the corrections should not affect these terms, as the solution should match on to $AdS_5\times S^5$ asymptotically. The first coefficient, 
$c_3(y)$, does not actually have any physical information because it can be eliminated by
going to the ``center of mass''  coordinate system or, in other words, we can always set the first moment of the matrix model eigenvalue distribution to zero
without loss of generality.
As we noted in footnote \ref{COM}, this is automatic for $SU(N)$ gauge group.

The Laurent expansion of $\omega_1$ in $z$ is given by
\be
\omega_1=\lambda\sum_{n=0}^{\infty}{\rho_{n}\over z^{n+1}}\, ,
\label{ome}
\ee
where $\rho_n$ are the moments of the eigenvalue distribution
introduced in (\ref{herm-mom}):
\ba
\rho_n= \int dx\, \rho(x)\, x^{n}.
\ea
We would like to express the functions that appear in the asymptotic expansion of the harmonic function $h_1$ (\ref{grahone}) in terms of the $\rho_n$'s, which contain the information about the eigenvalue distribution of the matrix model. 
This can be systematically computed  by plugging (\ref{ome}) into the definition of $h_1$ in (\ref{h1-h2}) and comparing the result with (\ref{grahone}). We get for the first few coefficients
\ba
c_4(y)&=&{8\Delta \rho_2\over \lambda} (1-2\cos{2y}),\cr
c_5(y)&=&-{64\Delta \rho_3 \over \lambda^{3/2}}\cos{2y} \sin{y},\cr
c_6(y)&=&{32\Delta \rho_4\over \lambda^2}(1-2\cos 2y+2\cos 4y)-\f{16\Delta \rho_2}{\lambda}(2-4\cos 2y+3\cos 4y),
\label{mapeoguay}
\ea
where 
$\Delta \rho_n\equiv\rho_n-\rho^0_{n}$ is the difference between the bubbling solution eigenvalue moments and the $AdS_5\times S^5$ moments (we have also encountered this difference of moments in the computation of the correlators in the gauge theory (\ref{Xi234})). In deriving these formulas, we have taken
\ba
\rho_0-\rho^0_0=0\,, \qquad
\rho_1-\rho^0_1=0 
.\ea
The first condition comes from fixing the radius of the two geometries, so that both are asymptotically $AdS_5\times S^5$ with the same radius of curvature.
The second one is the ``center of mass'' condition. One can also see from expanding Wigner's semi-circle law -- which controls the eigenvalue distribution of the $AdS_5\times S^5$  vacuum solution --  that $\rho^0_0=1$, 
$\rho^0_1=0$, $\rho_2^0=\lambda/4$, $\rho_3^0=0$, and $\rho_4^0=\lambda^2/8$.

The prescription in \cite{Skenderis:2006uy,Skenderis:2007yb} is to express the near boundary expansion of the metric and the other bulk fields in Fefferman-Graham form:
\ba
&&ds^2_5= \f{dZ^2}{Z^2}+\f{dX^idX^j}{Z^2}
\bigg(G_{(0)ij}(X)+Z^2 G_{(2)ij}(X)
\nn\\
&&\hspace{50mm}
+Z^4\left(G_{(4)ij}(X)+\log Z^2 h_{(4)ij}(X)\right)+\ldots\bigg)\, , \cr
&&\Phi^2(X,Z)=Z^2\left(\log Z^2 \Phi_{(0)}^2(X)+\tilde\Phi^2_{(0)}(X)
+\ldots\right)\, , \cr
&&\Phi^k(X,Z)=Z^{4-k}\Phi_{(0)}^k(X)+\ldots+Z^k\Phi^k_{(2k-4)}(X)+\ldots \qquad \mbox{ for } k>2\, ,
\label{FG}
\ea
where $Z$ is the Fefferman-Graham radial coordinate and $X^i$ are coordinates on the boundary. The first terms in these equations, $G_{(0)ij}$, $\Phi^2_{(0)}$, and $\Phi^k_{(0)}$, are the sources for the stress tensor and the chiral primary operators of the field theory, while 
$G_{(4)ij}, \tilde\Phi^2_{(0)}$, and $\Phi^k_{(2k-4)}$ are the normalizable modes of the fluctuations. For the bubbling 
solutions used in this paper, the non-normalizable modes that introduce sources vanish. 
 
Our first task is to perform a near boundary expansion of the bubbling supergravity solution. To do this, we introduce a radial coordinates $R$, which is related to the coordinate $x$ appearing in the bubbling solution (\ref{metric0}) by
\ba
x=\log\left(R+\sqrt{R^2+1}\right)\, . 
\ea
In these new coordinates, the $AdS_5\times S^5$ metric (\ref{vaccc}) is given by
\ba
ds^2=8c^2\left((R^2+1)ds^2_{AdS_2}+{dR^2\over R^2+1}+R^2ds^2_2
+dy^2+\cos^2{y}\, d\Omega_4\right).
\ea
In this coordinate system the conformal boundary is at $R\rightarrow \infty$, where the metric on it is that of $AdS_2\times S^2$.

 After some calculations, we get the asymptotic form of the deviations in (\ref{met-fluc}) up to order ${\cal O}\left(R^{-5}\right)$ terms:
\ba
\Delta_1&=&-{1\over32}\left(4c_4+\tan{y}\, \partial_y c_4 \right){1\over R^2}
-{1\over64}\left(5c_5+\tan{y}\, \partial_y c_5 \right){1\over R^3}\cr
&&-\f{1}{2048}\left[96 c_6+16\tan y \,\partial_y c_6-48 c_4^2-12 \tan y\,   \partial_y c_4^2\right.\cr
&&\hskip 1.8cm \left.-3\tan^2 y (\partial_y c_4)^2+64 c_4(2\cos2y-7)+32\tan y(\cos 2y+2)\partial_y c_4\right]\f{1}{R^4},
\cr
&&\cr
\Delta_2&=&-{1\over32}\left(4c_4+\tan{y}\,\partial_y c_4 \right){1\over R^2}
-{1\over64}\left(5c_5+\tan{y}\,\partial_y c_5 \right){1\over R^3}
\cr
&&-\f{1}{2048}\left[96 c_6+16\tan y \,\partial_y c_6-48 c_4^2-12 \tan y\,   \partial_y c_4^2\right.\cr
&&\hskip 1.8cm\left.-3\tan^2 y (\partial_y c_4)^2+64 c_4(2\cos2y+1)+32\tan y(\cos 2y-2)\partial_y c_4\right]\f{1}{R^4},
\cr
&& \cr
\Delta_4&=&{1\over32}\left(4c_4+\tan{y}\,\partial_y c_4 \right){1\over R^2}
+{1\over64}\left(5c_5+\tan{y}\,\partial_y c_5 \right){1\over R^3}\cr
&&+\f{1}{2048}\left[
96 c_6+16\tan y \,\partial_y c_6-16 c_4^2-4 \tan y\,   \partial_y c_4^2\right.\cr 
&&\hskip 1.8cm \left.-\tan^2 y (\partial_y c_4)^2+64 c_4\sec y \cos3y+32\tan y\cos 2y\, \partial_y c_4
\right]\f{1}{R^4},
\cr
&&\cr
\Delta_{\rho}&=&{1\over32}\left(4c_4+\tan{y}\,\partial_y c_4 \right){1\over R^2}
+{1\over64}\left(5c_5+\tan{y}\,\partial_y c_5 \right){1\over R^3}\cr
&&+\f{1}{2048}\left[
96 c_6+16\tan y \,\partial_y c_6-16 c_4^2-4 \tan y\,   \partial_y c_4^2\right.\cr
&&\left.\hskip 1.8cm-\tan^2 y (\partial_y c_4)^2-64 c_4 (2\cos 2y +5)-32 \tan y(\cos 2y+2)\, \partial_y c_4
\right]\f{1}{R^4}\, .
\cr &&\cr &&
\label{fluctuations}
\ea
We recall that the functions $c_m(y)$ are given in terms of the moments of the eigenvalue distribution by (\ref{mapeoguay}).
Therefore, we have written the deviations in terms of the matrix model data, the various moments of the eigenvalue distribution.


One finds similarly that the RR four-form deviation  (\ref{fourrrr}) is  given up to order ${\cal O}\left(R^{-5}\right)$ by
\ba
{F_y \over F^{(0)}_y  }&\equiv& 1+\Delta_{F_y}\cr &=&
1+{1\over 64}\left(16 c_4+8\tan{y}\,\partial_y c_4- \partial^2_y c_4\right){1\over R^2}+\f{1}{128}\left(20c_5+9\tan{y}\, \partial_y c_5-\partial^2_y c_5\right){1\over R^3}\cr
&&+\f{1}{256}\left[
24 c_6 +10\tan y \, \partial_y c_6-\partial_y^2 c_6-96\sin^2y\, c_4\right.\cr
&&\hskip 1.8cm\left.+10\sec y(\sin3y-\sin y)\, \partial_y c_4+\partial^2_y c_4(1-\sec y\cos 3y)
\right]\f{1}{R^4}\, .
\label{fluxexpand}
\ea

Here we note that the coordinate $R$ used in our expansions (\ref{fluctuations},\ref{fluxexpand}) is closely related to the Fefferman-Graham radial
coordinate $Z$ defined in (\ref{FG}). The relation is given by 
\ba
Z=\f{1}R- \f{1}{4R^3}+\Ocal(1/R^5).
\label{R-FG}
\ea
This relation allows us to use compute the correlators for the various local operators up to dimension $4$ by isolating the relevant term in the $1/R$ asymptotic expansion of the corresponding bulk fluctuation, even though $R$ is not the Fefferman-Graham coordinate.

Now that we have the explicit form of the deviations that we need to calculate our correlators, we expand the deviations 
 in  a basis of spherical harmonics of $S^5$. We decompose the metric and the RR 5-form into an $AdS_5\times S^5$ part and a perturbation
\ba
g_{MN}&=&g^{(0)}_{MN}+h_{MN},\cr
F_{5}&=&F^{(0)}_{5}+f_{5}
.\ea
In this subsection, we are interested   only in the fluctuations of the metric and of the 5-form on  $S^5$, which only contain scalar harmonics \cite{Kim:1985ez}\footnote{For the correlators computed in this paper, the fluctuations in the other supergravity fields do not contribute.}
\ba
h^a_a&=&\sum \pi^k Y^k,\cr
f_{abcde}&=&\sum b^k\Lambda^k\epsilon_{abcde}Y^k,\cr
h_{(ab)}&=&\sum \phi^k_{(s)} \nabla_{(a}\nabla_{b)}Y^k+\ldots,
\label{def-exp}
\ea
where $a,\,b,\ldots$ are indices on $S^5$, 
$\Lambda^k=-k(k+4)$ is the mass of the $k$-th scalar spherical harmonic,
 and $\epsilon$ is a volume form of $S^5$ with unit radius. The brackets on the indices   instruct us to take the symmetric and traceless combination. 
In the last line we omit terms involving
vector and tensor spherical harmonics that are irrelevant for us.
Later, for the computation of the stress tensor, we will also need 
\ba
h_{\mu\nu}=\sum h_{\mu\nu}^k Y^k\, , 
\ea
where $\mu$ and $\nu$ are indices on $AdS_5$. 

Since the circular Wilson loop preserves an $SO(5)$ subgroup of the R-symmetry group of ${\cal N}=4$ super Yang-Mills, so does the bubbling supergravity solution. Therefore, in performing the harmonic decomposition of the solution, only $SO(5)$ invariant spherical harmonics contribute. These spherical harmonics depend only  on the polar angle of the $S^5$, which we identify with the coordinate $y\in[-\pi/2,\pi/2]$ of the base. The metric on the sphere is given by $ds^2=dy^2+\cos^2{y}\, ds^2_{S^4}$, as can be seen from the  $AdS_5\times S^5$ vacuum solution (\ref{metric0}). 
The  $SO(5)$ invariant spherical harmonics are given in this coordinate system by\footnote{A brief review of $SO(5)$ invariant harmonics and of Gegenbauer polynomials can be found, for example, in the Appendix of \cite{Giombi:2006de}.} 
\ba
Y^J(y)=\Ncal_J\, c_J^{(2)}(\cos y)\, ,
\label{invharmo}
\ea
where $c_J^{(2)}(\cos y)$ are Gegenbauer polynomials and the normalization factors are chosen  as in (A.2) of \cite{Skenderis:2007yb}
and in (\ref{Y-norm})
\be
\int_{S^5}Y^JY^{J'}=\pi^3 z(J) \delta^{JJ'}\, ,\qquad z(J)=\f{1}{2^{J-1}(J+1)(J+2)}\, .
\ee
This fixes the normalization of the $SO(5)$ invariant spherical harmonics to 
\ba
{\cal N}_J=\sqrt{\frac{3 J!}{2^{J-1}(J+1) (J+3)!}}\, .
\ea
In particular, the explicit normalization for the harmonics we will need is given by
\be
{\cal N}_0=1\,,\qquad {\cal N}_1=\f{1}{4}\,,\qquad{\cal N}_2=\f{1}{2\sqrt{30}}\, ,\qquad {\cal N}_3=\f{1}{8\sqrt{10}} \,,\qquad {\cal N}_4=\f{1}{20\sqrt 7}\, .
\label{norm-sphhar}
\ee
For the reader's convenience, we list the explicit form of the first few harmonics:
\ba
Y^{0}&=&{\cal N}_{0}\, ,\cr
Y^{1}&=&{\cal N}_{1} 4\sin{y}\, ,\cr
Y^{2}&=&{\cal N}_{2}(-2+12\sin^2{y})\, ,\cr
Y^{3}&=&{\cal N}_{3}(-12\sin{y}+32\sin^3{y} )\, ,\cr
Y^{4}&=&{\cal N}_{4}(3-48\sin^2{y}+80\sin^4{y})\, .
\label{expl-harm}
\ea

More details about the expansion in fluctuations and about general properties of spherical harmonics are given in  \cite{Skenderis:2006uy,Skenderis:2007yb}.


\subsection{Chiral primary operators}

To compute the one-point functions for chiral primary operators ${\cal O}_J$  we  need appropriate combinations of the trace of the metric and of the RR four-form fluctuations, which are mass eigenstate of the Laplacian on the sphere \cite{Lee:1998bxa}
\be
s^k={1\over 20(k+2)}(\pi^k-10(k+4)b^{k}).
\ee
The first expansion in (\ref{def-exp}) can be inverted to find $\pi^k$ by using the orthogonality of the spherical harmonics 
\ba
\pi^k=\f{\int^{\pi/2}_{-\pi/2}dy\,  h^a_a\, Y^k \cos^4{y}}{
\int^{\pi/2}_{-\pi/2}dy\,  (Y^k)^2 \cos^4{y}}
\ea
where, in our case, $h^a_a=4\Delta_4+\Delta_{\rho}$, which follows from (\ref{met-fluc}). More precisely, we   need to pick the appropriate terms 
in the expansion (\ref{fluctuations}). 
The terms to compute depend on the dimension $J$ of the dual chiral primary operator ${\cal O}_J$. 
 For  ${\cal O}_2$ we have to select the  coefficient of the  $R^{-2}$ term in (\ref{fluctuations}), for ${\cal O}_3$  the coefficient of the term  $R^{-3}$  in (\ref{fluctuations}).
These are the same as the coefficients of $Z^2$ and $Z^3$, respectively,
due to (\ref{R-FG}).
The coefficient of the $Z^{k}$ term in the expansion of the quantity $A$ is usually denoted with the notation $[A]_k$. 
In general $[A]_k$ differs from the coefficient of $R^{-k}$.\footnote{
For the quantities relevant to our particular computations,
the differences  cancel out.
}
From the explicit expressions in (\ref{expl-harm}) and from the expansion for $\Delta_{4}$ and $\Delta_{\rho}$ in (\ref{fluctuations}) we get
\ba
\left[\pi^2\right]_2&=&\left[\hat{\pi}^2\right]_2={5\sqrt{30}\over\lambda}\Delta \rho_2\,,\cr
\left[\pi^3\right]_3&=&\left[\hat{\pi}^3\right]_3={20\sqrt{10}\over\lambda^{3/2}}\Delta \rho_3,\cr
\left[\pi^4\right]_4&=&
{\sqrt{7}\over 2\lambda^2}
\left(
-84\lambda\Delta \rho_2-45(\Delta \rho_2)^2+100\Delta \rho_4\right),
\ea
where hatted quantities in these formulas denote gauge invariant quantities at first order in the fluctuations. Note in particular that the scalars entering in the  dimensions 2 and 3 computations are automatically gauge invariant, whereas this is not the case for dimension 4 operators, as we will see presently.

Similarly, we can invert the second equation in (\ref{def-exp}) to get $b^k$:
\ba
\left[b^2\right]_2&=&[\hat{b}^2]_2=
\f{\int [\Delta_{F_y}]_2\left(-{1\over 12}\right)Y^2 \cos^4{y}}{
\int (Y^2)^2 \cos^4{y}}=-{\sqrt{30}\over 4\lambda}\Delta  \rho_2\,,\cr
\left[b^3\right]_3&=&[\hat{b}^3]_3=
\f{\int [\Delta_{F_y}]_2\left(-{1\over 21}\right)Y^3 \cos^4{y}}{
\int (Y^3)^2 \cos^4{y}}=-{2\sqrt{10}\over3\lambda^{3/2}}\Delta  \rho_3\,,\cr
\left[b^4\right]_4&=&
\f{\int [\Delta_{F_y}]_2\left(-{1\over 32}\right)Y^4 \cos^4{y}}{
\int (Y^4)^2 \cos^4{y}}={\sqrt{7}
\over 8\lambda^2}\left(9\lambda\Delta \rho_2-10 \Delta  \rho_4\right).
\ea
Again, dimensions 2 and 3 quantities are already gauge invariant, unlike $b^4$. 

Gauge invariant combinations for the $k=4$ case can be nonetheless easily formed using $\phi^4_{(s)}$, as explained in  \cite{Skenderis:2006uy,Skenderis:2007yb}. This coefficient can be obtained from the third expansion in (\ref{def-exp}), using some standard properties of spherical harmonics
\ba
\int D^{(a}D^{b)}h_{(ab)} Y^4=\phi^4_{(s)} 4\left(1+{\Lambda^4\over5}\right)\Lambda^4
\int (Y^4)^2,
\ea
so that, integrating by parts and using  (\ref{met-fluc}, \ref{fluctuations}) we get
\ba
[\phi^4_{(s)}]_4&=&{1\over 4\left(1+{\Lambda^4\over5}\right)\Lambda^4}
\f{\int D^{(a}D^{b)}[h_{(ab)}]_4Y^4}{ \int (Y^4)^2} \cr
&=&   {1\over 4\left(1+{\Lambda^4\over5}\right)\Lambda^4}
\f{\int \left([\Delta_\rho]_4\, \partial^2_y Y^4 - 4[\Delta_4]_4 \tan y \, \partial_y Y^4+\f{32}{5}[\Delta_\rho+4\Delta_4]_4Y^4\right) 
}{\int (Y^4)^2}
\cr &=&-{\sqrt{7}\Delta \rho_2\over 4\lambda}.
\ea
One can now form  the following gauge invariant combination of fluctuations \cite{Skenderis:2007yb}
\ba
\left[\hat{\pi}^4\right]_4&=&\left[\pi^4\right]_4-\Lambda^4 \left[\phi^4_{(s)}\right]_4
=\f{5\sqrt 7}{2\lambda^2}\left(20\Delta \rho_4-9(\Delta \rho_2)^2-20\lambda
\Delta \rho_2\right)\, ,
\cr
[\hat{b}^4]_4&=&  \left[b^4\right]_4-{1\over2}\left[\phi^4_{(s)}\right]_4=-{5\sqrt 7 \over 4\lambda^2}(\Delta \rho_4-\lambda\Delta \rho_2).
\ea
We have at this point all the ingredients to construct the gauge invariant mass eigenfunctions to linear order in the fluctuations:
\ba
\left[s^2\right]_2&=&{1\over80}(\left[\hat{\pi}^2\right]_2-60[\hat{b}^2]_2)={\sqrt{30}\over4\lambda}(\Delta \rho_2)\, ,\cr
\left[s^3\right]_3&=&{1\over100}(\left[\hat{\pi}^3\right]_3-70[\hat{b}^3]_3)={2\sqrt{10}\over3\lambda^{3/2}}(\Delta \rho_3)\, ,\cr
\left[s^4\right]_4&=&{1\over120}(\left[\hat{\pi}^4\right]_4-80[\hat{b}^4]_4)=\f{\sqrt 7}{16\lambda^2}
\left(20\Delta \rho_4-3(\Delta \rho_2)^2-20\lambda\Delta \rho_2\right).
\ea

Using  holographic renormalization  we are now able to extract the one-point functions of various local operators. Given the local five-dimensional supergravity action together with the associated counterterms, correlation functions can be computed by differentiating with respect to the non-normalizable modes($=$sources) of the bulk fields. In the gauge theory, we have computed the correlation functions of unit normalized operators  (\ref{unit}). The one-point functions of the unit normalized chiral primary operators ${\cal O}_2$ and ${\cal O}_3$ are given in terms of the supergravity fluctuations by  \cite{Skenderis:2007yb}
\ba
\left< {\cal O}_{2}(x)\right>_W&=&{N \over 2}{2\sqrt{8}\over3}[s^2]_2={N\over \lambda}{\sqrt{5 \over 3}}\Delta \rho_2,\cr
\left< {\cal O}_{3}(x)\right>_W&=&{3N \over 2\sqrt{6}}
[s^3]_3={N\over \lambda^{3/2}}{\sqrt{5 \over 3}}\Delta \rho_3\, ,
\label{lowdim}
\ea
 while the expectation value for the dimension 4 operator contains non-linear terms, as anticipated above,  and reads
\ba
\left< {\cal O}_{4}(x)\right>_W&=&
{N^2\over 2\pi^2}{4\sqrt{3}\over5}\left[2s^4+\f{2}{3z(4)}a_{422}
\left(s^2\right)^2\right]_4=\f{N}{\lambda^2}\f{\sqrt 7}{2}
\left(\Delta \rho_4-\lambda\Delta \rho_2
\right)\, , 
\label{dimfour}
\ea
with the triple overlap function $a_{422}$ being 
\be
a_{422}=\f{1}{\pi^3}\int Y^4 (Y^2)^2={\sqrt{7}\over 800}
\ee
and $z(4)=1/240$ as defined in (\ref{norm-sphhar}).
These are the final results from supergravity
for the one-point functions of low dimension chiral primary operators.

The non-trivial information about the correlator is in the function $\Xi_{R,J}$ defined in (\ref{WO-AdS})
\ba
{\left\langle
{\cal O}_J(x)
\right\rangle_{W}}
=\Xi_{R,J} Y(\theta).
\label{corr-AdS2S2-Xi2}
\ea
The dependence of the correlator on the choice of representative of the chiral primary multiplet factorizes, and it is captured by the spherical harmonic
function $Y(\theta)$, where $\theta^i$ determines the coupling of the scalars to the loop (\ref{W-def}).
Since the bubbling supergravity solution is $SO(5)$ invariant, the supergravity correlator computes the one-point function of a chiral primary operator which is $SO(5)$ invariant and can be constructed from the $SO(5)$ invariant spherical harmonics $Y=Y^J$ in (\ref{invharmo}). For the choice of scalar coupling we have made where  $\theta=(1,0,\dots,0)$, we have that 
\ba
Y^J(\pi/2)={\cal N}_J\f{(J+3)!}{6J!}=\sqrt{\frac{(J+2) (J+3)}{2^{J+1} 3}}.
\label{normaSO(5)}
\ea
Therefore, we can compute $\Xi_{R,J}$ by dividing (\ref{lowdim},\ref{dimfour}) by (\ref{normaSO(5)}) and obtain 
\ba
\Xi_{R,2}=\sqrt 2 \f{N}\lambda \Delta \rho_2,~~~
\Xi_{R,3}= 2\sqrt{\f{2}3}\f{N}{\lambda^{3/2}} \Delta \rho_3,~~~~
\Xi_{R,4}= 2\f{N}{\lambda^2} (\Delta \rho_4 -\lambda \Delta\rho_2).
\label{Xi234-sugra}
\ea
The agreement between 
the supergravity results (\ref{Xi234-sugra})
and the gauge theory computations (\ref{Xi234})
is then manifest!


\subsubsection{The small representation limit}

In comparing the gauge theory and supergravity results we have not used the explicit expressions for the moments $ \rho_n$. Here we evaluate them for the rank $k$ symmetric and antisymmetric representation respectively and for the specific case of $J=2$. For these small representations, the bulk computation of the correlator can be performed in terms of a probe  D3  and D5-brane respectively \cite{Giombi:2006de}. Even though the curvature gets large in the interior of the bubbling geometry, the curvature is small near the boundary, which is sufficient to compute these 
correlators.

From (\ref{lowdim}) above and from (\ref{sym2}) and (\ref{antisym2}) in Appendix \ref{mom-app}, we have (after transforming to $\mathbb{R}^4$) 
\be
\f{\langle W_R(\theta,a) {\cal O}_2(L)\rangle}{\langle W_R(\theta,a)\rangle}
=\f{4 a^2}{L^4}\f{N}{\lambda}\sqrt{\f{5}{3}}\Delta \rho_2=\left\{ \begin{array}{ll} 
\f{4a^2}{L^4}\sqrt{\f{5}{3}}\kappa\sqrt{1+\kappa^2} & ~~~ \mbox{symmetric case}\, , \\
& \\
\f{4a^2}{L^4}\sqrt{\f{5}{3}}\f{\sqrt\lambda}{6\pi}\sin^3\theta_k & ~~~\mbox{antisymmetric case}\, . \end{array}\right.
\label{sugradelta2}
\ee
On the other hand the probe D-brane computation in \cite{Giombi:2006de} gave
\be
\f{\langle W_R(\theta,a) {\cal O}_2(L)\rangle}{\langle W_R(\theta,a)\rangle}
=\f{a^2}{L^4} \, c_{S/A,2}\, Y^2\left(\f{\pi}{2}\right)\, ,
\ee
with 
\be
c_{S,2}=4\sqrt 2 \kappa\sqrt{1+\kappa^2}\, , \qquad c_{A,2}=\f{2\sqrt{2\lambda}}{3\pi}\sin^3\theta_k\, .
\ee
Putting everything together and recalling that $Y^2(\pi/2)=\sqrt{5/6}$ one finds exact agreement with (\ref{sugradelta2}). 


\subsection{Stress tensor}
\label{enmom-sect}

We now move on to the computation of the one-point function of the stress tensor.
As explained in  \cite{Skenderis:2006uy,Skenderis:2007yb}, we need to compute $\left( 1+{1\over3}\pi^0 \right)g^0_{\mu\nu}+h^0_{\mu\nu}$. Here $h^0_{\mu\nu}$ are the zero-modes of the metric perturbation on $AdS_5$ (recall that $Y^0=1$)
\be
h^0_{\mu\nu}=\f{8}{3\pi}\int^{\pi/2}_{-\pi/2}dy\,  h_{\mu\nu}\, \cos^4{y}\, , 
\ee
which explicitly read
\ba
h^0_{\mu\nu}dx^\mu dx^\nu&=&\left(R^2+1\right)
\left(\f{3\Delta \rho_2\left(5\Delta \rho_2-2\lambda\right)}{8\lambda^2R^4}+{\cal O}(R^{-5}) \right)
ds^2_{AdS_2}\cr
&&+ \left(-\f{\Delta \rho_2\left(15\Delta \rho_2-8\lambda\right)}
{24\lambda^2R^4}+{\cal O}(R^{-5})\right) {dR^2\over R^2+1} \cr
&&+R^2\left(\f{\Delta \rho_2\left(45\Delta \rho_2+14\lambda\right)}{24\lambda^2R^4}+{\cal O}(R^{-5})\right)
ds^2_{S^2}\, ,
\ea
while $\pi^0$ is the zero mode of the trace of  the metric perturbation $h^a_a$
\ba
\pi^0=\f{8}{3\pi}\int^{\pi/2}_{-\pi/2}dy\,  h_{a}^a\, \cos^4{y}=-\f{25(\Delta
\rho_2)^2}{8\lambda^2 R^4}+{\cal O}(R^{-5})\, .
\ea
Then, the modified metric
\ba
ds^2=\left( 1+{1\over3}\pi^0 \right)g^0_{\mu\nu}+h^0_{\mu\nu} dx^\mu dx^\nu
\label{modified-10d-met}
\ea 
is expanded in $1/R$ as
\ba
ds^2&=&(R^2+1) 
\left(1+\f{\Delta \rho_2(10\Delta\rho_2-9\lambda)}{12\lambda^2R^4}+{\cal O}(R^{-5})\right)
ds^2_{AdS_2}\cr
&&+\left(1+\f{\Delta \rho_2(-5\Delta \rho_2+\lambda)}{3\lambda^2R^4}+{\cal O}(R^{-5})
 \right){dR^2\over R^2+1}\cr
&&+R^2
\left(1+\f{\Delta \rho_2(10\Delta \rho_2+7\lambda)}{12\lambda^2R^4}+{\cal O}(R^{-5}) \right)
ds^2_2 \cr
&\equiv&
(R^2+1) \left(1+ {p_1\over R^4}\right)
ds^2_{AdS_2}+ \left(1+ {p_2\over R^4}\right)
{dR^2\over R^2+1}+R^2
\left(1+ {p_3\over R^4}\right)ds^2_2 
\, .\ea

We now introduce the following near boundary coordinate
\be
R={1\over z}
\left(
1-{1\over4}z^2+{p_2\over 8}z^4
\right)\, ,
\ee
so that  the metric becomes
\ba
ds^2&=&
\left({1\over z^2}+{1\over2}+\left({1\over 16}+p_1+{p_2\over 4} \right)z^2 \right)
ds^2_{AdS_2}+{dz^2\over z^2}\cr
&&+\left({1\over z^2}-{1\over2}+\left({1\over 16}+{p_2\over4}+p_3 \right)z^2 \right)
ds^2_{S^2}\, .
\label{metfeff}
\ea
Note  that $z$ is the near boundary coordinate for the modified
metric (\ref{modified-10d-met})
and thus differs form Fefferman-Graham $Z$ introduced in (\ref{R-FG}).

From \cite{Skenderis:2007yb}, the stress tensor correlator is given by\footnote{Note that $g_{(n)}$ denotes the coefficient of the $z^n$ term in the expansion of  (\ref{metfeff}), {\it i.e.} \newline
$ds^2=\f{dz^2}{z^2}+\f{1}{z^2}(g_{(0)ij}+\ldots+z^ng_{(n)ij})dx^idx^j$.}
\ba
\left< T_{ij}\right>&=&{N^2\over2\pi^2}
\left(g_{(4)ij}-{2\over9} \left([\hat{s^2}]_2\right)^2g_{(0)ij}
\right.\cr&&\hskip 1.8cm \left.
+{1\over8}[\Tr g^2_{(2)}-\left(\Tr g_{(2)} \right)^2]g_{(0)ij}
-{1\over2}\left( g^2_{(2)}\right)_{ij}+{1\over4}g_{(2)ij}\Tr g_{(2)}\right)\, ,
\ea
where the $g_{(k)ij}$'s are the 
 analogues of the quantities in (\ref{FG})
for the metric 
(\ref{modified-10d-met}).
 
Plugging  (\ref{metfeff}) into the expression above
and using $[s^2]_2={\sqrt{30}\over4\lambda}\Delta \rho_2$, which appears in
the dimension 2 chiral primary calculation, we get
\ba
\langle T_{ij}(x)\rangle_W dx^i dx^j
=-\f{N^2}{3\pi^2 \lambda}\Delta \rho_2
(ds^2_{AdS_2}-ds^2_{S^2})+ \f{N^2}{32\pi^2}(ds^2_{AdS_2}+ds^2_{S^2}).
\ea
This precisely agrees with
the gauge theory computation  (\ref{hW2}) 
\ba
h_W&=& 
-\f{N}{3\sqrt 2 \pi^2} \Xi_{R,2}=-\f{N^2}{3\pi^2 \lambda}\Delta \rho_2\, ,
\label{hW2-sugra}
\ea
including the conformal anomaly contribution
(\ref{T-ads})
\ba
\langle T_{ij}(x)\rangle_W dx^i dx^j
=h_W(ds^2_{AdS_2}-ds^2_{S^2})+ \f{N^2}{32\pi^2}(ds^2_{AdS_2}+ds^2_{S^2})\, .
\label{T-ads-sugra}
\ea


\subsection*{Acknowledgments}

We are happy to thank David Berenstein, Evgeny  Buchbinder, Nadav Drukker, Kostas Skenderis, and Marika Taylor for helpful discussions and correspondence.
Research at Perimeter Institute for Theoretical Physics is supported in part by the Government of Canada through NSERC and by the Province of Ontario through MRI. J.G. and S.M. also acknowledge further support from an NSERC Discovery Grant.  T.O. and D.T.  are partly supported by the NSF grants 
PHY05-51164 and PHY-04-56556. D.T. is also supported by the Department of Energy under Contract DE-FG02-91ER40618.
\vfill\eject

\appendix


\section{Weyl transforms between  boundary metrics}
\label{sec-conformalfactor}

In this appendix we discuss the two Weyl transformations
relating $\R^4$ and $AdS_2\times S^2$, which we have used in section \ref{loop-CPO}. The first transformation is relevant for the circular loop computation, while the second one for the straight line.

Let us parametrize $\R^4$
using two sets of polar coordinates so that
\be
ds^2_{\mathbb{R}^4}=dr^2+r^2d\psi^2+dL^2+L^2 d\phi^2. 
\ee
These coordinates are relevant for a circular loop, which we take to be defined 
by $r=a$ and $L=0$. 
By making the following change of coordinates
\be
\tilde r^2={(r^2+L^2-a^2)^2+4a^2L^2\over 4a^2}
={a^2\over(\cosh\rho-\cos\theta)^2}\,,\qquad
r=\tilde r\sinh\rho\,,\qquad
L=\tilde r\sin\theta\,,
\ee
we find the metric
\be 
ds^2_{\mathbb{R}^4}=\tilde r^2\left(d\rho^2+\sinh^2\rho\,d\psi^2+
d\theta^2+\sin^2\theta\,d\phi^2\right)\,,
\label{conff}
\ee
which is conformal to  $AdS_2\times S^2$ in global coordinates. Note that the conformal factor $\tilde{r}$ is that in (\ref{tilda}) and that the loop, which was located at $r=a,\, L=0$ in $\R^4$, gets mapped to the conformal boundary of $AdS_2\times S^2$, namely the boundary of the Poincar\'e disk.

Now, under the conformal transformation (\ref{conff}) a dimension $J$ operator ${\cal O}_J$ transforms as follows: ${\cal O}_J\rightarrow \tilde{r}^{-J} {\cal O}_J$. This proves the relation between the form of the correlator in $\R^4$ (\ref{corr-coef}) and the one in $AdS_2\times S^2$ 
(\ref{WO-AdS}).

The metric for $\R^4$ can also be written as
\ba
ds^2_{\mathbb{R}^4}=dt^2+dl^2+l^2 ds^2_{S^2}.
\ea
We place the straight line at $l=0$.
In this case the Weyl transformation to $AdS_2\times S^2$
is simple:
\ba
ds^2_{\mathbb{R}^4}=l^2 ds^2_{AdS\times S^4}\, ,
\ea
where
\ba
ds^2_{AdS_2\times S^2}=\f{dt^2+dl^2}{l^2}+ds^2_{S^2}
\ea
involves the $AdS_2$ metric in Poincar\'e coordinates.
The operators transform as $\Ocal_J\ra l^{-J} \Ocal_J$
when going from $AdS_2\times S^2$ to $\R^4$,
thus proving (\ref{corr-coef-line}).


\section{Relating the stress tensor to a chiral primary via a
GL twist}
\label{GL-sec}

Here we rederive the relation (\ref{relacioentre})
between the correlator of the Wilson line with the stress tensor
and the correlator of the Wilson line with the dimension two chiral primary ${\cal O}_2$
from a different point of view.

The basic observation is  that the supersymmetric Wilson line is closed with respect to the BRST charge of the Geometric Langlands (GL) twist  \cite{Kapustin:2006pk}.\footnote{More precisely, the GL twists form a 1-parameter family of twists, where the parameter, $t$, is the projective coordinate on $\mathbb{C}P^1$. The Wilson line is closed with respect to the $t=i$ twist.} Since, by definition of a topological field theory, the twisted stress tensor $T'_{\mu\nu}$ is BRST exact, it follows that the expectation value of $T'_{\mu\nu}$ in the presence of the Wilson line is zero:
\be
\langle W_R(\theta,a) \,T'_{\mu\nu}\rangle = \langle W_R(\theta,a)\, \{Q_{GL}, V_{\mu\nu}\}\rangle=-\langle V_{\mu\nu}\,\{Q_{GL},W_R(\theta,a)\}\rangle =0\, .
\ee
We can then compute the difference between the stress tensors of the twisted and untwisted theories and consider its correlator with $W_R(\theta,a)$. This will turn out to give the wanted relation with the correlator of ${\cal O}_2$.

We do not need to consider the kinetic term of the gauge field
since this is an R-symmetry singlet  not affected by the twist. 
We also ignore fermions at first.
Before the twist the action in a generic curved background is
\be
S=\f{1}{g^2_{YM}}\int d^4x \sqrt{g}\, \Tr\left(D_\mu \phi^i D^\mu \phi^i +\f{R}{6}\phi^i\phi^i -\f{1}{2}[\phi^i,\phi^j]^2\right)\, ,
\label{S-untwisted}
\ee
where, as before, $i=1,\ldots,6$.
We now identify an $SO(4)\subset SO(6)$ with the Lorentz group, so that $i=\mu,5,6$, and define $\sigma=\f{1}{\sqrt 2}(\phi^5+i\phi^6)$. The GL twisted action is then given by (see equations (3.46 - 3.48) in \cite{Kapustin:2006pk})
\ba
S'&=&\f{1}{g^2_{YM}}\int d^4x \sqrt{g}\, \Tr\Big(D_\mu \phi_\nu D^\mu \phi^\nu +R_{\mu\nu}\phi^\mu\phi^\nu -\f{1}{2}[\phi_\mu,\phi_\nu]^2\cr &&\hskip 4cm+2{D}_\mu \sigma {D}^\mu \bar\sigma-2[\phi_\mu,\sigma][\phi^\mu,\bar\sigma]  +[\sigma,\bar\sigma]^2\Big)\, .
\ea
The covariant derivatives $D_\mu$ contain both the gauge and metric connections.
In flat space $S=S'$.

Let us now compute the stress tensor by taking the variation of the action with respect to the metric and setting in the end $g_{\mu\nu}=\delta_{\mu\nu}$. One finds 
\ba
&&T_{\mu\nu}-T'_{\mu\nu}=\f{2}{\sqrt g}\f{\delta}{\delta g^{\mu\nu}}(S-S')\Big|_{g_{\mu\nu}=\delta_{\mu\nu}}\cr && \hskip 1cm =\f{2}{g^2_{YM}}\,\Tr\Big(
\f{1}{6}\delta_{\mu\nu}{D}^2(\phi^i\phi^i)-\f{1}{6}{D}_\mu {D}_\nu(\phi^i\phi^i)+\phi_{(\mu} {D}^2\phi_{\nu)}+{D}_\rho(\phi_{(\mu}{D}_{\nu)}\phi_\rho)\cr
&& \hskip 3cm \left.-{D}_\rho(\phi_\rho {D}_{(\mu}\phi_{\nu)}) -\f{1}{2}\delta_{\mu\nu} {D}_\rho {D}_\sigma (\phi_\rho\phi_\sigma)-\f{1}{2}{D}^2(\phi_\mu\phi_\nu)+{D}_\rho {D}_{(\mu}(\phi_{\nu)}\phi_\rho)\right.\cr
&& \hskip 3cm +[\phi_\mu,\phi_\rho][\phi_\nu,\phi_\rho]+2[\phi_{(\mu},\sigma][\phi_{\nu)},\bar\sigma]
\Big)\, ,
\label{T}
\ea
where we have used the following formulas
\ba
\delta \Gamma_{\mu\nu}^\rho &=& \f{1}{2}g^{\rho\sigma}({D}_\mu \delta g_{\nu\sigma}+{D}_\nu \delta g_{\mu\sigma}-{D}_\sigma \delta g_{\mu\nu})\,,\cr
\delta R&=& R_{\mu\nu}\delta g^{\mu\nu}+g_{\mu\nu} {D}^2 \delta g^{\mu\nu}-{D}_\mu {D}_\nu \delta g^{\mu\nu}\, ,\cr
\delta R_{\rho\sigma}&=& \f{1}{2}g_{\mu\nu} {D}_{(\rho} {D}_{\sigma)} \delta g^{\mu\nu} +\f{1}{2}g_{\rho\mu}g_{\sigma\nu}{D}^2 \delta g^{\mu\nu}-g_{\mu (\rho} {D}_\nu {D}_{\sigma)}\delta g^{\mu\nu}\, ,
\label{GR-ident}
\ea
and integrated by parts.
Imposing  the equations of motion
\be
D^2 \phi_\mu=-[\phi_\rho,[\phi_\mu,\phi_\rho]]-[\sigma,[\phi_\mu,\bar\sigma]]-[\bar\sigma,[\phi_\mu,\sigma]]
\ee
we can eliminate the quartic terms in (\ref{T}) and arrive at the final expression
\ba
T_{\mu\nu}-T'_{\mu\nu}&=&\f{2}{g^2_{YM}}\,\Tr\left(\f{1}{6}\delta_{\mu\nu}{D}^2(\phi^i\phi^i)-\f{1}{6}{D}_\mu {D}_\nu(\phi^i\phi^i)+{D}_\rho(\phi_{(\mu}{D}_{\nu)}\phi_\rho)\right.\cr && \hskip .5cm\left.-{D}_\rho(\phi_\rho {D}_{(\mu}\phi_{\nu)})-\f{1}{2}\delta_{\mu\nu} {D}_\rho {D}_\sigma (\phi_\rho\phi_\sigma)-\f{1}{2}{D}^2(\phi_\mu\phi_\nu)+{D}_{(\mu}{D}_\rho(\phi_{\nu)}\phi_\rho)\right)\, .\cr &&
\label{T1}
\ea
Note that on-shell $\partial^\mu(T_{\mu\nu}-T'_{\mu\nu})=0$, as it should be.

Let us take now the Wilson loop to be a line along the $\mu=1$ direction and consider the correlation function with the stress tensor. By $SO(5)$ symmetry we can say that $\langle W_R(\theta,a)\Tr (\phi^\alpha\phi^\beta)\rangle=\langle W_R(\theta,a) \Tr (\phi^2\phi^2)\rangle\delta^{\alpha\beta}$ and $\langle W_R(\theta,a)\Tr (\phi^1\phi^\alpha)\rangle=0$ where $\alpha=2,\ldots,6$. Moreover, derivatives in the $\mu=1$ direction vanish because of translational invariance and $\langle W_R(\theta,a)\Tr (\phi^\alpha \partial_\mu \phi^\beta) \rangle $ is non zero only if $\alpha=\beta$. Putting everything together we have (here $a=2,3,4$)
\ba
&&\langle W_R(\theta,a) T_{11}\rangle=-\f{2}{3g^2_{YM}}
\partial_a\partial_a\left\langle W_R(\theta,a)\Tr\left(\phi^1\phi^1-\phi^2\phi^2\right)\right\rangle\,, \qquad \langle W_R(\theta,a) T_{1a}\rangle=0 \cr 
&&
 \langle W_R(\theta,a)T_{ab}\rangle=-\f{1}{3g^2_{YM}}
(\partial_a\partial_b-\delta_{ab}\p_c\p_c)\left\langle W_R(\theta,a) \Tr
\left(\phi^1\phi^1-\phi^2\phi^2\right)\right\rangle 
\label{Tfin}
\ea
and it is also immediate to realize that $\langle W_R(\theta,a) T_\mu^{\; \mu}\rangle=0$.

The operator $\Tr\left(\phi^1\phi^1-\phi^2\phi^2\right)$
 is a chiral primary of dimension two.
The dependence ($\propto 1/l^2$) of
 $\langle W_R(\theta,a) \Tr\left(\phi^1\phi^1-\phi^2\phi^2\right)\rangle$
is consistent with the functional form (\ref{T-line}) of
$\langle W_R(\theta,a) T_{\mu\nu}\rangle$.
One then finds that
\ba
h_W&=& -\f{4}{3 g_{YM}^2}\left\langle\Tr\left( \phi^1\phi^1-\phi^2\phi^2\right)\right
\rangle_W
=
-\f{N}{3\sqrt 2 \pi^2} \Xi_{R,2}.
\label{hW}
\ea
Here we have used $Y(\theta)=\f{1}{{\sqrt 2}}( \theta^1\theta^1-\theta^2\theta^2)$ evaluated at $\theta=(1,0,\ldots,0)$.

We see that the bosonic contributions
have already reproduced the full result (\ref{relacioentre}).
Thus it should be possible to show that the fermionic contributions
sum up to zero, though we do not perform this computation here.


\section{Moments 
in the small representation limit}
\label{mom-app}

In this appendix we compute the explicit expressions for the second moments $\langle \xi^2\rangle$ of the hermitian eigenvalue distributions in the cases of rank $k$ symmetric and antisymmetric representations. Wilson loops transforming in these representations are described by D3 and D5 probe branes, respectively, having $k$ units of string charge dissolved in their worldvolumes \cite{Drukker:2005kx,Yamaguchi:2006tq,Gomis:2006sb}.


\subsection{Symmetric case}

Let us call the eigenvalues $\xi_i$ and label them in increasing order, $\xi_1<\ldots <\xi_N$. 
Without the Wilson loop insertion, the eigenvalues are distributed on the interval $[-\sqrt\lambda,\sqrt\lambda]$ and satisfy Wigner's semi-circle law
derived from the saddle point equations
\ba
-\f{4N}\lambda \xi^{(0)}_i+\sum_{j\neq i}\f{2}{\xi^{(0)}_i-\xi^{(0)}_j}&=&0\qquad
\hbox{~for~all~} i=1,\ldots,N.\label{noloop}
\ea
As is well-known, the last eigenvalue is $\xi_N^{(0)}=\sqrt\lambda$
and the resolvent of the matrix model  is given by $\omega_0(\zeta)= g_{YM}^2\sum_i 1/(\zeta-\xi_i^{(0)})=2\zeta-2\sqrt{\zeta^2-\lambda}$.

Inserting in the path integral a Wilson loop in the rank $k$ symmetric representation consists in moving the last eigenvalue $\xi_N$ a distance $k$ away from the interval.
The saddle point equations become then
\ba
-\f{4N}\lambda \xi_i+\sum_{j\neq i, j\neq N}\f{2}{\xi_i-\xi_j}+\f{2}{\xi_i-\xi_N}&=&0\qquad
\hbox{~for~} i=1,\ldots,N-1, \label{Sk-SPE1} \\
-\f{4N}\lambda \xi_N+k+\sum_{j=1}^{N-1}\f{2}{\xi_N-\xi_j}&=&0\, . \label{Sk-SPEII}
\ea
We make the ansatz that
\ba
\xi_i=\xi_i^{(0)}+\delta \xi_i,\qquad\delta \xi_i=\Ocal(1/N)\qquad\hbox{for } i=1,\ldots,N-1\, ,
\ea
while the shift for the $\xi_N$ is large as remarked above.
Then (\ref{Sk-SPEII}) implies that
\ba
\xi_N=\sqrt\lambda \sqrt{1+\kappa^2}+\Ocal(1/N),
\ea
where $\kappa\equiv\sqrt\lambda k/4N$.
We want to compute
\ba
\langle \xi^2\rangle-\langle \xi^2\rangle_0
&=&
\oo N \sum_{i=1}^{N-1} (\xi_i^{(0)}+\delta \xi_i)^2
+\oo N \xi_N^2-\oo N \sum_{i=1}^N \xi_i^{(0)}{}^2
\nn\\
&=&\f{2}N \sum_{i=1}^{N-1} \xi_i^{(0)}\delta \xi_i +\oo N \xi_N^2-\oo N \xi_N^{(0)}{}^2
+\Ocal(1/N^2).
\ea

By taking the difference between
the two saddle point equations (\ref{noloop}) and (\ref{Sk-SPE1}) for $i=1,\ldots ,N-1$, we get
 \ba
 -\f{4N}\lambda \delta \xi_i -2\sum_{j\neq i,j\neq N}\f{\delta \xi_i-\delta \xi_j}{(\xi_i^{(0)}-\xi_j^{(0)})^2}-\f{2}{\xi_i^{(0)}-\xi_N^{(0)}}+\f{2}{\xi_i^{(0)}-\xi_N}
=\Ocal(1/N).
 \ea
We multiply this equation  by $\xi_i^{(0)}$ and sum over $i$
from 1 to $N-1$.
By playing with the dummy indices, one can show that
\ba
-2\sum_{i=1}^{N-1} \xi_i^{(0)}\sum_{j\neq i, j\neq N}
\f{\delta \xi_i-\delta \xi_j}{(\xi_i^{(0)}-\xi_j^{(0)})^2}
&=&-2\mathop{\sum_{i=1}^{N-1}\sum_{j=1}^{N-1}}_{i\neq j}
\f{\delta \xi_i}{\xi_i^{(0)}-\xi_j^{(0)}}
\cr &=&-\f{4N}\lambda\sum_{i=1}^{N-1} \xi_i^{(0)}\delta \xi_i+\Ocal(1).
\label{Sk-sum}
\ea
In the second line we used the saddle point equation for $\xi_i^{(0)}$.
Thus
\ba
-\f{8N}\lambda \sum_{i=1}^{N-1}\xi_i^{(0)}\delta \xi_i
+\sum_{i=1}^{N-1}\xi^{(0)}_i\left(-\f{2}{\xi_i^{(0)}-\xi_N^{(0)}}+\f{2}{\xi_i^{(0)}-\xi_N}\right)
=\Ocal(1/N)\, , 
\ea
from which follows
\ba
\f{ 2}{N}\sum_{i=1}^{N-1}\xi_i^{(0)}\delta \xi_i&=&\oo{2N}(-\xi_N\omega_0(\xi_N)
+\xi^{(0)}_N\omega_0(\xi_N^{(0)}))+\Ocal(1/N^2).
\ea
By collecting everything, we get
\ba
\langle \xi^2\rangle-\langle \xi^2\rangle_0=\f{\lambda}N\kappa
\sqrt{1+\kappa^2}+\Ocal(1/N^2).
\label{sym2}
\ea


\subsection{Antisymmetric case}

The effect of inserting a Wilson loop in the antisymmetric representation is to create a hole in the $[-\sqrt\lambda,\sqrt\lambda]$ interval so that the distribution splits into two groups with $k$ and $N-k$ eigenvalues.\footnote{Note the difference between the symmetric and antisymmetric representation: the former can have arbitrary rank $k$, consistently with the fact that we can move $\xi_N$ arbitrarily far away from the interval, whereas the latter must have $k\le N$, with the hole confined inside the interval. } The shift is ${\cal O}(1/N)$ for all of them and the saddle point equations read
\ba
-\f{4N}\lambda \xi_i+\sum_{j\neq i}\f{2}{\xi_i-\xi_j}&=&0\qquad
\hbox{~for~} i=1,\ldots,N-k,\cr
-\f{4N}\lambda \xi_i+1+\sum_{j\neq i}\f{2}{\xi_i-\xi_j}&=&0\qquad
\hbox{~for~} i=N-k+1,\ldots,N\,. \label{Ak-SPEII}
\ea
Subtracting (\ref{noloop}) from these expressions one gets
\ba
-\f{4N}\lambda \delta \xi_i+2\sum_{j\neq i}\f{\delta \xi_j-\delta \xi_i}{(\xi_i^{(0)}-\xi_j^{(0)})^2}&=&{\cal O}\left(1/N\right)\qquad
\hbox{~for~} i=1,\ldots,N-k,\cr
-\f{4N}\lambda \delta \xi_i+1+2\sum_{j\neq i}\f{\delta \xi_j-\delta \xi_i}{(\xi_i^{(0)}-\xi_j^{(0)})^2}&=&{\cal O}\left(1/N\right)\qquad
\hbox{~for~} i=N-k+1,\ldots,N\, .\cr&&\label{Ak-SPEV}
\ea
In this case the expression for the second moment reads
\ba
\langle \xi^2\rangle-\langle \xi^2\rangle_0=\f{2}{N}\sum_{i=1}^N \xi_i^{(0)}\delta \xi_i+{\cal O}\left(1/N^2\right)\, .
\ea
Multiplying (\ref{Ak-SPEV}) by $\xi_i^{(0)}$, summing over the respective ranges of $i$, and finally summing the two equations one has
\ba
-\f{4N}\lambda \sum_{i=1}^N \xi_i^{(0)}\delta \xi_i+2\sum_{i=1}^N \xi_i^{(0)}\sum_{j\neq i}\f{\delta \xi_j-\delta \xi_i}{(\xi_i^{(0)}-\xi_j^{(0)})^2}+\sum_{i=N-k+1}^N
\xi_i^{(0)}={\cal O}\left(1/N\right)\, .
\ea
Using (\ref{Sk-sum}) and defining $\xi=\sqrt\lambda\cos\theta$, one has up to orders $\mathcal{O}\left(1/N^{2}\right)$
\ba
\f{2}{N}\sum_{i=1}^N \xi_i^{(0)}\delta \xi_i&=&\f{g^2_{YM}}{4N}\sum_{i=N-k+1}^N
\xi_i^{(0)}=\f{g^2_{YM}}{4}\f{2}{\pi\lambda}\int_{\xi_k}^{\sqrt\lambda}d\xi\sqrt{\lambda-\xi^2}\cr
&=&\f{\lambda^{3/2}}{2\pi N}\int_{0}^{\theta_k}d\theta \sin^2\theta \cos\theta=\f{\lambda^{3/2}}{6\pi N}\sin^3\theta_k\, .
\label{antisym2}
\ea
\vfill\eject

\bibliography{correlators}  

\end{document}